\documentclass[aps,pra,reprint,groupedaddress]{revtex4-1}
\usepackage[UKenglish]{babel}
\usepackage[]{graphicx, amsfonts, amsmath, amssymb, amstext, latexsym, float, color, hyperref}
\usepackage{tikz}
\usetikzlibrary{decorations.pathreplacing,calligraphy,circuits.logic.US}
\usepackage{bm}
\newcommand{\be}{\begin{equation}}
\newcommand{\ee}{\end{equation}}
\newcommand{\ket}[1]{|#1\rangle}
\newcommand{\bra}[1]{\langle#1|}
\newcommand{\eat}[1]{}

\begin{document}

\title{Quantum-enhanced quickest change detection of transmission loss}

\author{Saikat Guha$^{(1,2)}$, Tiju Cherian John$^{(2,3)}$, Zihao Gong$^{(1)}$ and Prithwish Basu$^{(4)}$}
\affiliation{$^{(1)}$Department of Electrical and Computer Engineering, University of Maryland, College Park MD}
\affiliation{$^{(2)}$Wyant College of Optical Sciences, University of Arizona, Tucson AZ}
\affiliation{$^{(3)}$Department of Mathematics, BITS Pilani, K K Birla Goa Campus, Goa,  India
}
\affiliation{$^{(4)}$RTX BBN Technologies, Cambridge MA}

\begin{abstract}
Augmenting a train of bright phase-modulated laser-light pulses of a coherent communications system with infinitesimally small quantum photons per pulse---entangled across several time bins---prepared by splitting squeezed light in a temporal-mode interferometer can dramatically enhance a homodyne receiver’s ability to detect a sudden change in the channel loss, by up to a factor that is the inverse of the pre-change loss, without affecting the communications rate. We discuss the quantum limit of quickest change detection, and the problem of joint communications and change detection that our study opens up.
\end{abstract}
\maketitle

{\em Introduction}---Active fiber monitoring methods, e.g., optical time-domain reflectometry (OTDR)~\cite{Shim2012} have been investigated at length, for detecting and preventing wiretapping intrusion or to nullify the significance of information tapped in optical communication channels~\cite{Iqbal2011}. Physical-layer security~\cite{Fok2011} and attack detection~\cite{Medard2002} methods have been studied extensively for all-optical networks. In recent years, drawing upon the security proofs of weak-coherent-state based quantum key distribution (QKD), tap detection schemes have been devised that rely on insertion of pilot tones into classical communications signaling~\cite{Gong2020}. Concurrently, it has been long known that squeezed light can be more sensitive than classical laser light (coherent-state) as a probe for estimating an unknown loss in optical propagation~\cite{Monras2007-jg, Gong2023}, for distinguishing between two prior-known values of loss, e.g., for optical reading from a disk~\cite{Pirandola2011-xo}, and for communicating classical information over a lossy optical channel~\cite{Shapiro1979}.

In this Letter, we propose a readily-realizable scheme for bolstering an existing coherent-detection optical communications system to detect a sudden increase in channel loss. The simplest instance of our protocol augments bright laser-light phase-modulated pulses of a codeword by tiny amounts of squeezing. An advanced version of our protocol sprinkles weak multi-mode continuous-variable entanglement, generated by splitting a squeezed vacuum pulse in a temporal multi-mode splitter~\cite{Cui2023}, across a block of bright modulated laser-light pulses. Using the theory of {\em quickest change detection} (QCD), we show that a vanishingly small quantum energy, added to the bright classical pulses, using our entanglement-based method, leads to a sharp decrease in the latency with which the receiver can identify a sudden increase in the channel loss for a given time-to-false-alarm. The maximum drop in latency afforded by our quantum-augmentation scheme is by a factor that is the inverse of the pre-change loss. 

{\em Quickest change detection}---The simplest version of the change-point-detection problem goes as follows~\cite{Page1957}. Alice is handing i.i.d. copies of random variables $X_i$, $1 \le i \le k$, to Bob, where $k$ is a running discrete time index. For $i \in \left\{1, \ldots, n_c-1\right\}$, Alice picks each copy of $X_i$ from distribution $P_1(x)$, whereas for $i \in \left\{n_c, \ldots, k\right\}$, she picks $X_i$ i.i.d. from $P_2(x)$. We assume that distributions $P_1(x)$ and $P_2(x)$ are known to Bob. Bob's task is to detect the change point $n_c$ as quickly and as accurately as possible. The cumulative sum (CUSUM) algorithm proceeds as follows. Bob calculates the CUSUM $S[k] = \sum_{n=1}^k l[n]$ of the log-likelihood ratio $l[n] = \log[P_2(X_n)/P_1(X_n)]$ and the decision function $G[k] = \max_{1 \le n_c \le k}\sum_{n=n_c}^k l[n]$. The {\em maximum likelihood} (ML) estimate of the change-point, ${\hat n_c} = {\rm argmax}_{1 \le n_c \le k}\sum_{n=n_c}^k l[n]$ $=  {\rm argmin}_{1 \le n_c \le k} S[n_c - 1]$ is the time following the current minimum of the CUSUM. The first time $G[k] > h$, Bob declares a change indeed occurred and records $k \to n_d$, as the time when a change was detected. This is called the generalized likelihood ratio test. The latency of change-point detection $\tau = n_d - n_c$ quantifies how quickly Bob catches the change (at a level of confidence quantified by $h$) after the change occurred. The minimum time to false alarm $\gamma(h)$ is the expected value of $n_d$ when no change actually occurred (i.e., $n_c = \infty$). The aforesaid CUSUM algorithm achieves the optimum value of the worst-case latency~\cite{Lorden1971}, 
\begin{equation}
\tau_{\rm min} \sim \frac {\log \gamma(h)}{S(P_2 || P_1)},
\label{eq:taumin_CUSUM}
\end{equation}
where $S(P_2 || P_1) = \int P_2(x)\log [P_2(x)/P_1(x)] dx$ is the Kullback-Leibler divergence, or {\em classical relative entropy} (CRE) between probability distributions $P_2$ and $P_1$. The quantum limits of change-point detection have been recently studied in theory~\cite{Sentis2017, Sentis2018, Fanizza2022-tu, Fanizza2023-ch, John2025-bz} and an experiment that detected a change in the state of emitted photons~\cite{Yu2018}.

{\em Baseline problem setup}---Alice is using a pulsed laser transmitter to emit a steady stream of coherent-state pulses $|\alpha\rangle^{\otimes k}$, with $\alpha \in {\mathbb R}$ and mean photon number per pulse, $\alpha^2$, which upon traversal through a lossy channel, are each detected by homodyne detection at the receiver. We assume that the transmissivity of the channel---possibly including any out-coupling losses at the transmitter and detection inefficiency of the receiver---is $\eta_1$, and known to Bob. At an unknown time $n_c$, $1 \le n_c \le k$, the channel's transmissivity drops to $\eta_2 = \eta_1 \eta_{\rm tap}$, e.g., due to the appearance of a malicious wiretap. Let us assume that Bob also knows $\eta_{\rm tap}$, and hence $\eta_2$, which is simple to relax. The homodyne detection receiver produces a $k$-vector ${\boldsymbol X} = \left\{X_1, X_2, \ldots, X_k\right\}$ of i.i.d. Gaussian-distributed random variables $X_i \sim {\mathcal N}(\sqrt{\eta_j}\,\alpha, 1/4)$, where $j=1$ pre-change ($i < n_c$) and $j=2$ post-change ($i \ge n_c$). The $1/4$ is the quantum limited variance of local-oscillator shot-noise limited homodyne detection~\footnote{Incorporating excess noise, such as stemming from electronic noise or local oscillator intensity fluctuations causing excess noise due to imperfect common mode rejection ratio (CMRR) of homodyne detection, results in qualitative changes to results presented herein, but are not important for the main message of the Letter.}. We will denote the mean photon number per pulse for the classical baseline case described above, $\alpha^2 \equiv N + N_a$. $N_a \ll N$ is a small-signal contribution to the mean energy per pulse, insignificant for the classical case, but included for a fair comparison with the quantum-augmented cases discussed below where the $N_a$ photons are quantum. The classical relative entropy between the post-change distribution $P_2 \sim {\cal N}\sqrt{\eta_2}\,\alpha, 1/4)$ and the pre-change distribution $P_1 \sim {\cal N}\sqrt{\eta_1}\,\alpha, 1/4)$ of the homodyne output, $S(P_2 || P_1) \equiv S^{(0)}$, is given by:
\begin{equation}
S^{(0)} = 2(N+N_a)(\sqrt{\eta_2}-\sqrt{\eta_1})^2 \equiv S_c(N,N_a,\eta_1,\eta_2),
\label{eq:S0}
\end{equation}
and $\gamma(h)$, the average run length (ARL), i.e., the expected value of the first value of $k$ when $G[k]$ exceeds $h$, is given by the solution of Fredholm integral equations as described by Page~\cite{Page1957}, a numerical approximation of which was given by Goel and Wu~\cite{Goel1971}. Appendix~\ref{app:ARL} describes our ARL computation for numerical examples.
\begin{figure}
\centering
\includegraphics[width=\columnwidth]{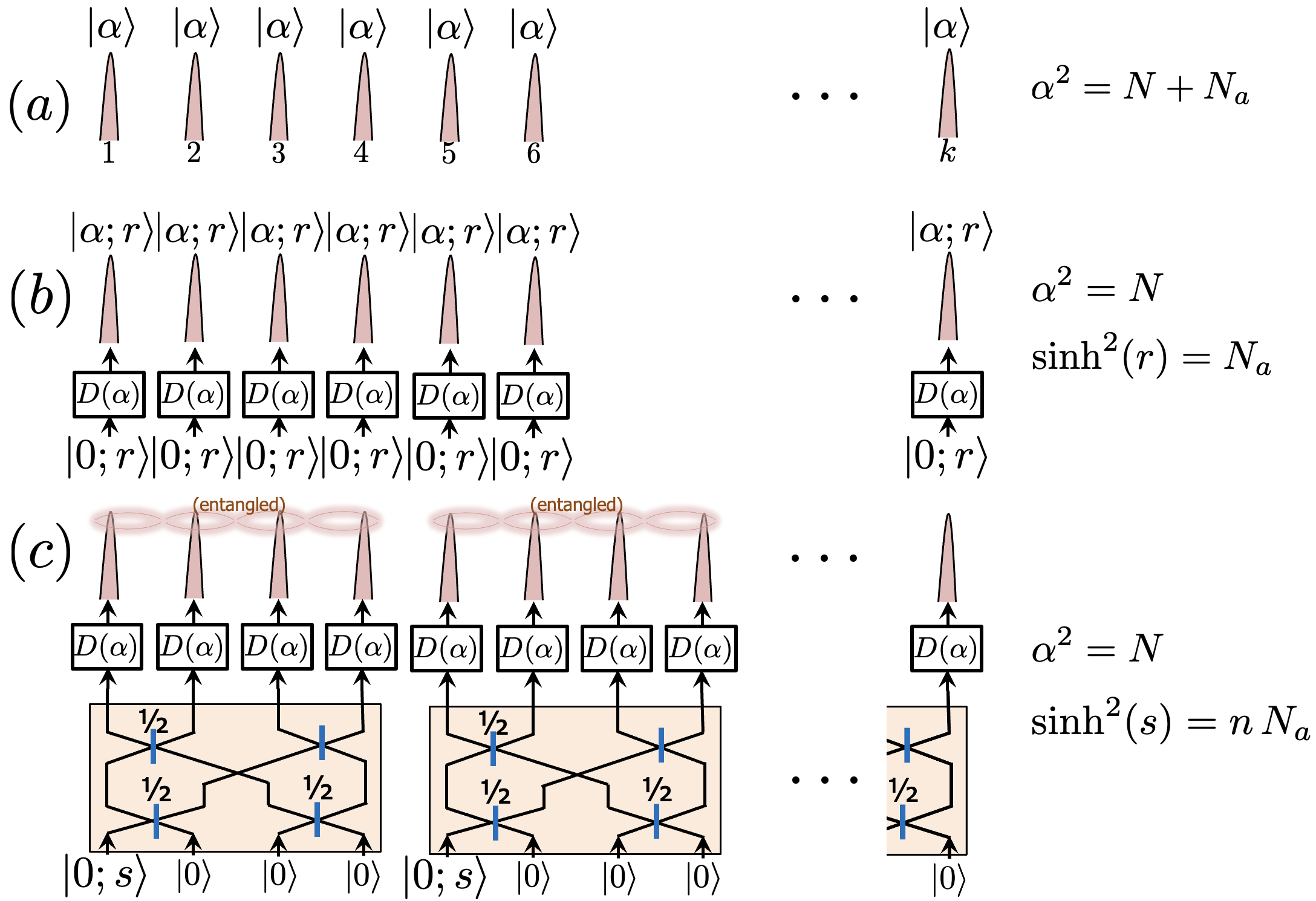}
\caption{(a) Identical laser-light pulses of amplitude $\alpha$, $\alpha \in {\mathbb R}$ (i.e., tensor product of coherent states $|\alpha\rangle^{\otimes k}$), each of mean photon number $|\alpha|^2 \equiv N+N_a$, $N_a \ll N$. (b) Identical squeezing-augmented laser-light pulses, a.k.a. displaced squeezed states of light, $|\alpha; r\rangle^{\otimes k}$, where $|\alpha|^2 \equiv N \gg N_a \equiv {\rm sinh}^2(r)$, i.e., the energy attributable to squeezing ($N_a$) is much less compared to the photon energy in the coherent amplitude ($N$). (c) Blocks of $n$ laser-light pulses $|\alpha\rangle^{\otimes n}$, each with mean photon number $N = |\alpha|^2$ are augmented by a continuous variable (CV) entangled state, generated by splitting a squeezed-vacuum pulse of mean photon number $nN_a$ in an $n$-mode equal splitter, such that the photon energy attributable to quantum augmentation, per pulse, is $N_a \ll N$.}
\label{fig:system_figure}
\end{figure}

{\em Squeezing-augmented transmitter}---Let us assume that the transmitted pulses are in a displaced squeezed state $|\alpha; r\rangle = D(\alpha)|0; r\rangle$, with $\alpha^2 = N$ photon equivalent of classical coherent energy and ${\rm sinh}^2(r) = N_a \ll N$ quantum photon energy per pulse. Assuming real-quadrature squeezing, and that at the output of the lossy channel the homodyne-detection receiver's local oscillator is aligned with the squeezed quadrature, the homodyne output is a $k$-vector ${\boldsymbol X} = \left\{X_1, X_2, \ldots, X_k\right\}$ of i.i.d. Gaussian-distributed random variables $X_i \sim {\mathcal N}(\sqrt{\eta_j}\,\alpha, v_j)$, with $v_j = [\eta_j e^{-2r} + (1-\eta_j)]/4$, where $j=1$ pre-change ($i < n_c$) and $j=2$ post-change ($i \ge n_c$). The classical relative entropy $S(P_2 || P_1) \equiv S^{(1)}$ is given by,
\begin{equation}
S^{(1)} = \frac{(v_2/v_1) - \log(v_2/v_1) - 1}{2} + \frac{N(\sqrt{\eta_2}-\sqrt{\eta_1})^2}{2 v_1}.
\label{eq:S1}
\end{equation}
\begin{figure}
\centering
\includegraphics[width=\columnwidth]{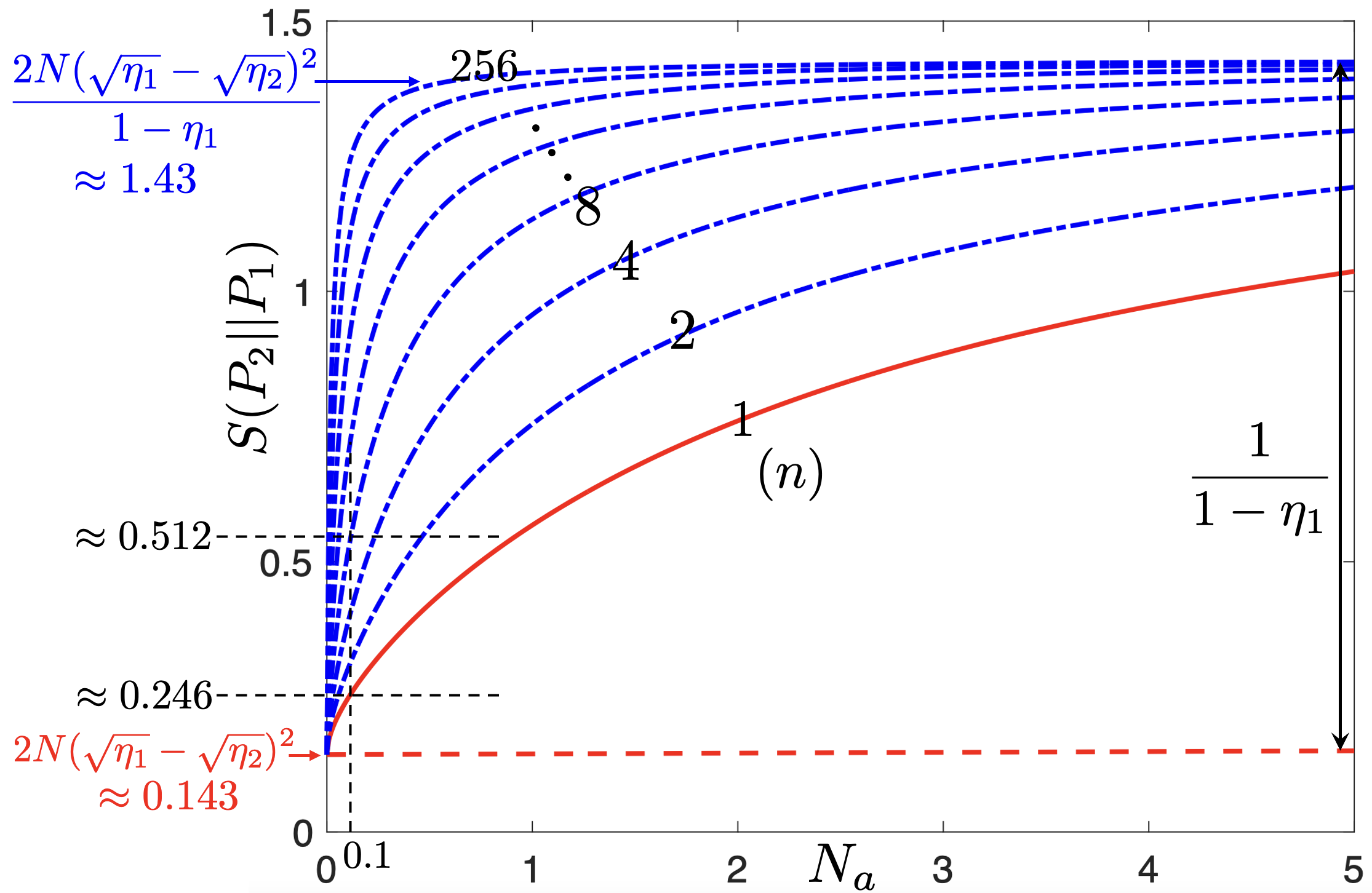}
\caption{Relative entropy $S(P_2||P_1)$ between the post-change and pre-change distributions at the homodyne-detection receiver's output, per detected pulse, as a function of $N_a \in [0, 5]$ photons, for $N = 100$ photons, $\eta_1 = 0.9$ and $\eta_2 = 0.85$, for the: (a) classical baseline (red-dashed), (b) squeezing-augmented (red solid), and (c) entanglement-augmented (blue dash-dotted plots, $n = 2, 4, \ldots, 256$) transmitters.}
\label{fig:CRE_improvement}
\end{figure}
We plot $S^{(1)}$ (red solid plot) and $S^{(0)}$ (red dashed plot) in Fig.~\ref{fig:CRE_improvement} for $N=100$ photons, $\eta_1 = 0.9$ and $\eta_2 = 0.85$, as a function of $N_a \in [0, 5]$ photons. The first thing we note is that $\partial S^{(1)}/\partial N_a = \infty$ at $N_a = 0$ (see Appendix~\ref{sec:derivative}), which is in sharp contrast to $\partial S^{(0)}/\partial N_a = 2(\sqrt{\eta_2}-\sqrt{\eta_1})^2$. For $N_a = 0.1$ (which corresponds to $r \approx 0.31$, i.e., $10\log_{10}(e^{2r}) \approx 2.7$ dB of squeezing), squeezing-augmentation affords $\sim 1.72$ fold increase in $S(P_2||P_1)$, which would translate to roughly $1.72$ fold reduction in the latency of detecting a sudden change in the channel's initial transmissivity of $0.9$, if we were to pick the CUSUM thresholds $h_0$ and $h_1$ for the two cases (classical-baseline and squeezing-augmented), such that the ARL to false alarm in the event of no change, $\gamma = \gamma_0(h_0) = \gamma_1(h_1)$ is held the same for both cases. 
\begin{figure}
\centering
\includegraphics[width=0.75\columnwidth]{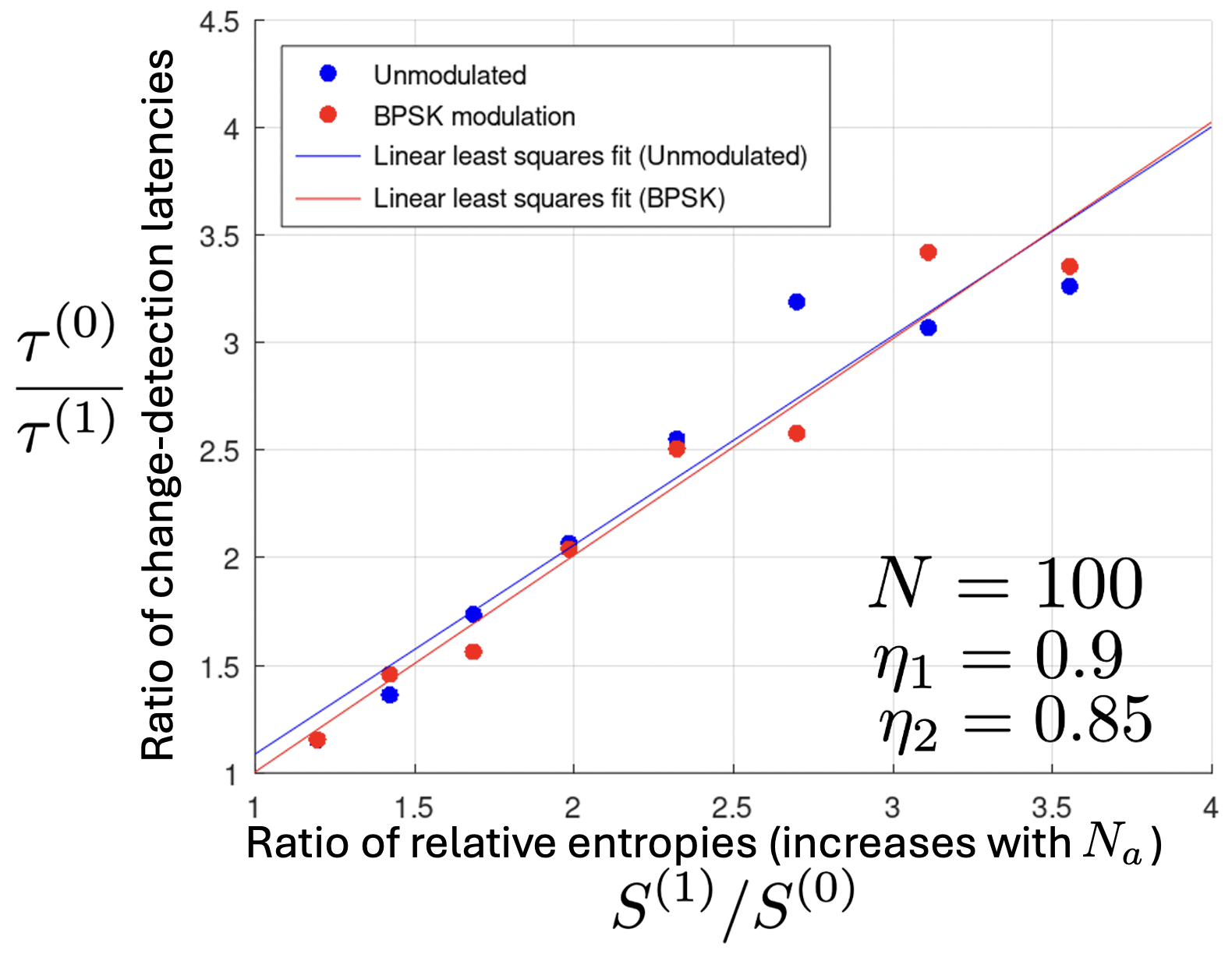}
\caption{Monte-Carlo plot of factor-of-improvement of detection latencies, $\tau^{(0)}/\tau^{(1)}$, as a function of $S^{(1)}/S^{(0)}$ (for ARL, $\gamma = 2$ million held constant for both) shows a $1:1$ correlation, as expected~\cite{Page1957}. The parameters chosen for the simulation: $N=100$, $N_a \in (0.01, 1]$, $\eta_1 = 0.9$, $\eta_2 = 0.85$, $n_c = 1000$, and all the simulations are run until $k = 5000$.}
\label{fig:CUSUM_squeezed}
\end{figure}
In Fig.~\ref{fig:CUSUM_squeezed}, we show a Monte-Carlo scatter plot (see blue dots) of the ratios of the simulated detection latencies, $\tau_0/\tau_1$, for the classical (unaugmented) versus the squeezing-augmented cases, plotted as a function of the relative-entropy ratios $S^{(1)}/S^{(0)}$, for $N=100$, $N_a \in (0.01, 1]$, $\eta_1 = 0.9$, and $\eta_2 = 0.85$. The true change occurs at time step $n_c = 1000$, and all simulations are run until $k = 5000$ timesteps. The CUSUM thresholds $h_0$ and $h_1$ are chosen such that the ARL, $\gamma = 2$ million (time steps) for both cases. As expected from Eq.~\eqref{eq:taumin_CUSUM}, we see that the slope of the scatter plot is roughly $1$. Next, we note from Eq.~\ref{eq:S1} that $\lim_{N_a \to \infty} S^{(1)} = ((1-\eta_2)/(1-\eta_1))+ 2N(\sqrt{\eta_2}-\sqrt{\eta_1})^2/(1-\eta_1)$, which shows that the maximum factor of improvement possible to be had is roughly $1/(1-\eta_1)$ (assuming the contribution of the $((1-\eta_2)/(1-\eta_1))$ term is small compared to the term proportional to $N$). In the example for Fig.~\ref{fig:CRE_improvement}, $1/(1-\eta_1) = 10$. 

{\em Entanglement-augmented transmitter}---Next, we consider the entanglement-augmented transmission scheme sketched in Fig.~\ref{fig:system_figure}(c). Blocks of $n$ coherent state pulses, $|\alpha\rangle^{\otimes n}$, each with mean photon number $N = |\alpha|^2$, are augmented by a continuous variable entangled state generated by splitting a squeezed-vacuum pulse $|0;s\rangle$ of mean photon number $nN_a = {\rm sinh}^2(s)$ in an $n$-mode equal splitter, such as the Hadamard unitary~\cite{Guha2011} realized using the temporal-mode Green Machine~\cite{Cui2023} (See Appendix~\ref{app:entangled_transmitter_schematic}). So, just like for the squeezing-augmented transmission discussed earlier, the photon energy attributable to quantum augmentation, per pulse, is $N_a$, and as before we will assume $N_a \ll N$. After lossy transmission (with pre-change and post-change transmissivities $\eta_1$ and $\eta_2$ respectively), each pulse is detected via a homodyne detection receiver. Each successive $n$-block of the homodyne output ${\boldsymbol X}_1^n \equiv \left\{X_1, \ldots, X_n\right\}$ is a correlated uniform-symmetric Gaussian random vector with means and covariances (see Appendix~\ref{app:ent_augmented} for derivations) given by: 
\begin{eqnarray}
E[X_l] &=& \sqrt{\eta_j} \,\alpha, \,\forall l,\label{eq:mean}\\
E[\Delta_l^2] &=& \eta_j \left(\frac{e^{-2r}+n-1}{4n}\right) + \frac{1-\eta_j}{4}, \forall l, \,{\text{and}}\label{eq:cov1}\\ 
E[\Delta_l \Delta_m] &=& \eta_j \left(\frac{e^{-2r} - 1}{4n}\right), \forall (l, m)\label{eq:cov2},
\end{eqnarray}
where $\Delta_k = X_k-E[X_k]$, $k=1,2,\dots,n$, $j=1$ pre-change ($i < n_c$), and $j=2$ post-change ($i \ge n_c$). The relative entropy, calculated on a per-pulse basis for fair comparison with the previous cases, $S^{(n)} \equiv S(P_2({\boldsymbol X}_1^n) || P_1({\boldsymbol X}_1^n))/n$, where $S(P_2({\boldsymbol X}_1^n) || P_1({\boldsymbol X}_1^n)) = \frac12 \left[{\rm Tr}(K_1^{-1}K_2) - \ln \frac{|K_2|}{|K_1|} -n\right] + \frac12 (u_2-u_1)K_1^{-1}(u_2-u_1)^{\rm T}$, where $K_j$ are the covariance matrices of the Gaussian distributions $P_j({\boldsymbol x})=P[{\boldsymbol X}_1^n = {\boldsymbol x}]$, $j = 1$ and $2$, $P_j({\boldsymbol x}) = \frac{1}{(2\pi)^{n/2}|K_j|^{\frac12}} {\rm exp}\left[-\frac12 ({\boldsymbol x}-u_j)K_j^{-1}({\boldsymbol x}-u_j)^{\rm T}\right]$, with all $n$ entries of the mean vector $u_j$ being $E[X_l]$ from Eq.~\eqref{eq:mean} while the diagonal and off-diagonal entries of the $n$-by-$n$ covariance matrix $K_j$ are given by $E[X_l^2]$ and $E[X_lX_m]$ from Eqs.~\eqref{eq:cov1} and~\eqref{eq:cov2} respectively.

\begin{figure}
\centering
\includegraphics[width=\columnwidth]{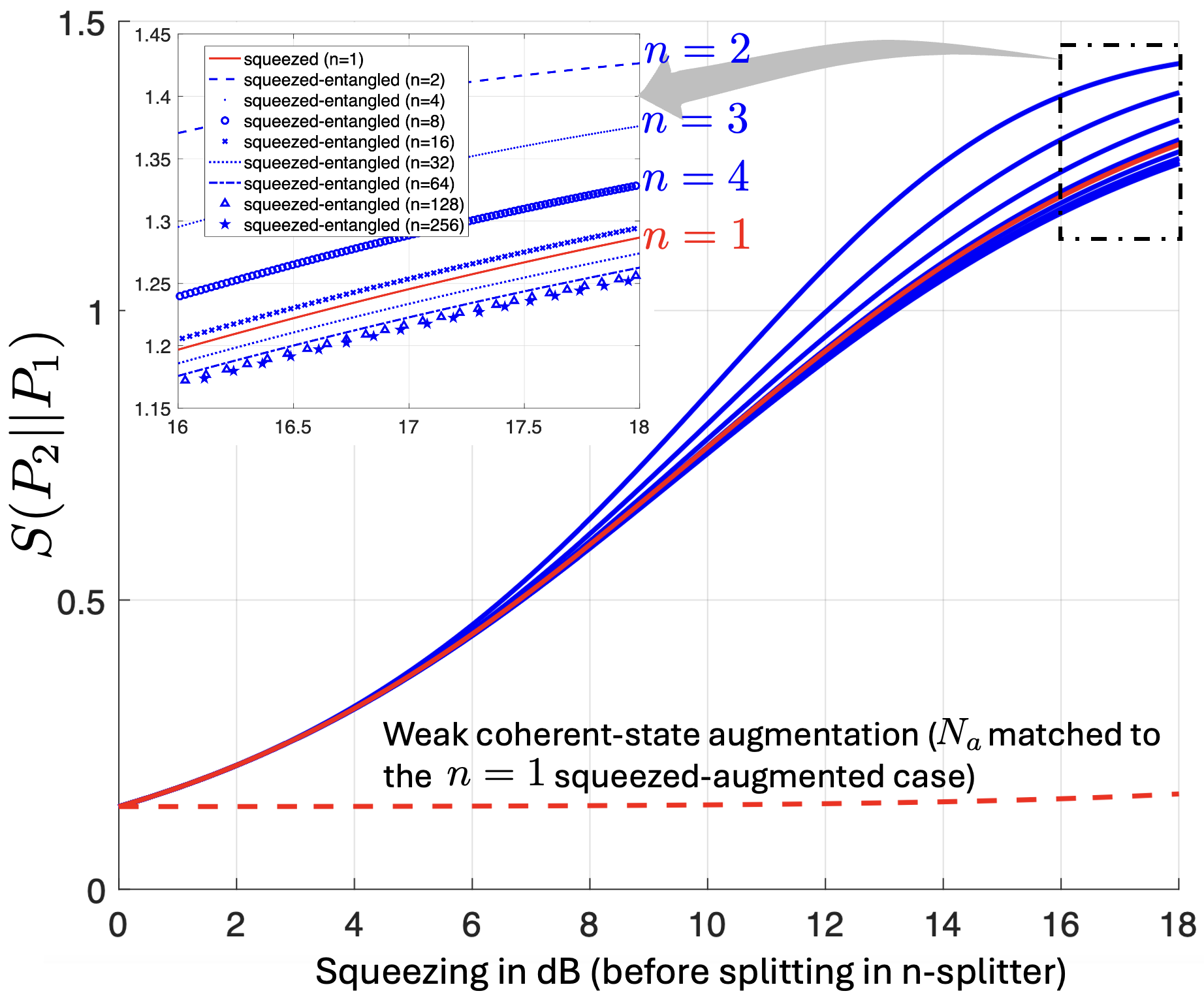}
\caption{The plot of $S(P_2||P_1)$ with respect to $10\log_{10}(e^{2s})$, the dB-squeezing of the seed squeezed-light source, used to generate the $n$-mode entanglement state. With $s$ held fixed, the $n=2$ case is seen to outperform all others.}
\label{fig:fixed_squeezing}
\end{figure}
We plot $S^{(n)}$ (blue dash-dotted plots) for $n = 2, 4, 8, \ldots, 256$ in Fig.~\ref{fig:CRE_improvement} for $N=100$ photons, $\eta_1 = 0.9$ and $\eta_2 = 0.85$, as a function of $N_a \in [0, 5]$ photons. Not only is $\partial S^{(n)}/\partial N_a = \infty$ at $N_a = 0$, $\forall n \ge 1$, the increase of $S^{(n)}$ from the classical value $S^{(0)} = 2N(\sqrt{\eta_2}-\sqrt{\eta_1})^2$ to the maximum afforded by squeezing-augmented transmission becomes progressively sharper as $n$ increases. In other words, adding an infinitesimally small amount of quantum-augmented photons to bright coherent-state pulses can yield up to a factor of $1/(1-\eta_1)$ reduction in the change-detection latency. One practical caveat to this result is that for a given $N_a$, as the entanglement block-length $n$ increases, the needed dB-squeezing of the initial squeezer, $10\log_{10}(e^{2s})$ increases. In Fig.~\ref{fig:fixed_squeezing}, we see that if we keep this initial squeezing held fixed, the $n=2$ entanglement-augmented transmitter outperforms all others. One might argue however that one could generate the $n$-mode entanglement using multiple weak squeezers and multi-mode mixing, much like the output state of Gaussian Boson Sampling~\cite{Hamilton2017}, rather than using a single strong squeezer. Whether such a strategy helps achieve performance at or above $S^{(n)}$ is left open. We present numerical results of simulated latencies $\tau^{(n)}$ for the entanglement-augmented case ($n \ge 2$) in Section~\ref{sec:endmatter}(b).

{\em Joint communications and change-detection}---Instead of the small-signal $N_a$ quantum photons riding on $N$-photon fixed-amplitude coherent states $|\alpha\rangle$, $|\alpha|^2 = N$, we will now consider the exact same quantum augmentations but on a train of binary-phase-shift-keying (BPSK) modulated coherent states $|\alpha\rangle$ or $|-\alpha\rangle$. With no augmentation, homodyne detection yields a BPSK Additive White Gaussian Noise (AWGN) channel. The Shannon capacity of this binary-discrete-input continuous-output channel is given by $C_{\text{BPSK-AWGN}} = \int_{-\infty}^{\infty} P(x|\xi=1)\log_2[2P(x|\xi=1)/(P(x|\xi=1)+P(x|\xi=-1))]dx = 1 - \int_{-\infty}^{\infty} P(x|\xi=1)\log_2(1 + e^{-2x/\sigma^2})dx$ bits per pulse, where $P(x|\xi=1) = (1/\sqrt{2\pi} \sigma)e^{-(x-1)^2/2\sigma^2}$, $\sigma^2 = 1/(4\eta N)$, and $\xi = \pm 1$ correspond to the phase of the received coherent state at the output of the lossy channel of transmissivity $\eta$. There is a vast literature on capacity-approaching error correction codes for this channel, of the convolutional-turbo families, using soft-information-passing decoders~\cite{Huettinger2006}. Since the homodyne-detection receiver's output can be re-utilized for quickest change-detection, when the signal-to-noise ratio (SNR) of the AWGN channel $4\eta N$ is high, such that the symbol-error probability is low at the receiver (hence the phase of the transmitted coherent state is correctly detected) almost all the conclusions on quantum-enhanced change detection presented thus far should prevail without affecting the communications rate, since $N_a \ll N$ means the AWGN channel's SNR is minimally altered. To demonstrate this empirically, in our simulated CUSUM results shown in Fig.~\ref{fig:CUSUM_squeezed}, we included the case of BPSK modulation with squeezing augmentation (red data points). Here, the CUSUM algorithm uses the pre- and post-change distributions to be $(P(x|\xi=1)+P(x|\xi=-1))/2$ with $\sigma^2$ taken as $1/(4\eta_1 N)$ versus $1/(4\eta_2 N)$ respectively, instead of taking the pre- and post-change distributions to be $P(x|\xi=1)$ as done for the unmodulated case. It is hard to discern any difference visually between the scatter plots of the blue (unmodulated) and the red (modulated) data points. The main point of the above paragraph was to show that the quantum augmentation is quite resilient to the laser communications capacity. The choice of the specific (BPSK) modulation format was just for ease of calculations, and not central to the argument.

{\em Quantum limit of quickest change detection}---Let us consider the baseline case of unmodulated coherent state transmission, $|\alpha \rangle^{\otimes k}$, with $\alpha^2 = N+N_a$. The {\em quantum relative entropy} (QRE) between the post-change state $|\sqrt{\eta_2}\alpha\rangle$ and the pre-change state $|\sqrt{\eta_1}\alpha\rangle$ at the channel output---which gives the ultimate limit of change detection latency~\cite{Fanizza2022-tu, John2025-bz}---equals infinity, since both are pure states. This suggests the existence of a receiver such that the classical relative entropy (CRE), $S(P_2 || P_1) = \infty$, where $P_1$ and $P_2$ are the pre- and post-change probability distributions at the output of that receiver; which would thereby achieve instantaneous change detection, per the CUSUM theory. The receiver that achieves this happens to be the Kennedy Receiver~\cite{Kennedy1973}, which applies a phase-space `nulling' displacement $D(-\sqrt{\eta_1}\alpha)$ to each received pulse followed by single-photon detection. The pre-change state, after the displacement, incident on the detector is therefore vacuum, $|0\rangle$, and the post-change state is $|(\sqrt{\eta_2}-\sqrt{\eta_1})\alpha \rangle$. The binary-valued output distributions, where `click' $\equiv 0$ and `no-click' $\equiv 1$, are given by: $P_1[0]=1$, $P_1[1]=0$; and $P_2[0] = \exp(-(\sqrt{\eta_2}-\sqrt{\eta_1})^2(N+N_a))$, $P_2[1] = 1 - P_2[0]$. The CRE, $S(P_2||P_1) = P_2[0]\log[P_2[0]/P_1[0]] + P_2[1]\log[P_2[1]/P_1[1]]$ clearly equals infinity (due to the second term) and hence equals the QRE.

\begin{figure}
\centering
\includegraphics[width=\columnwidth]{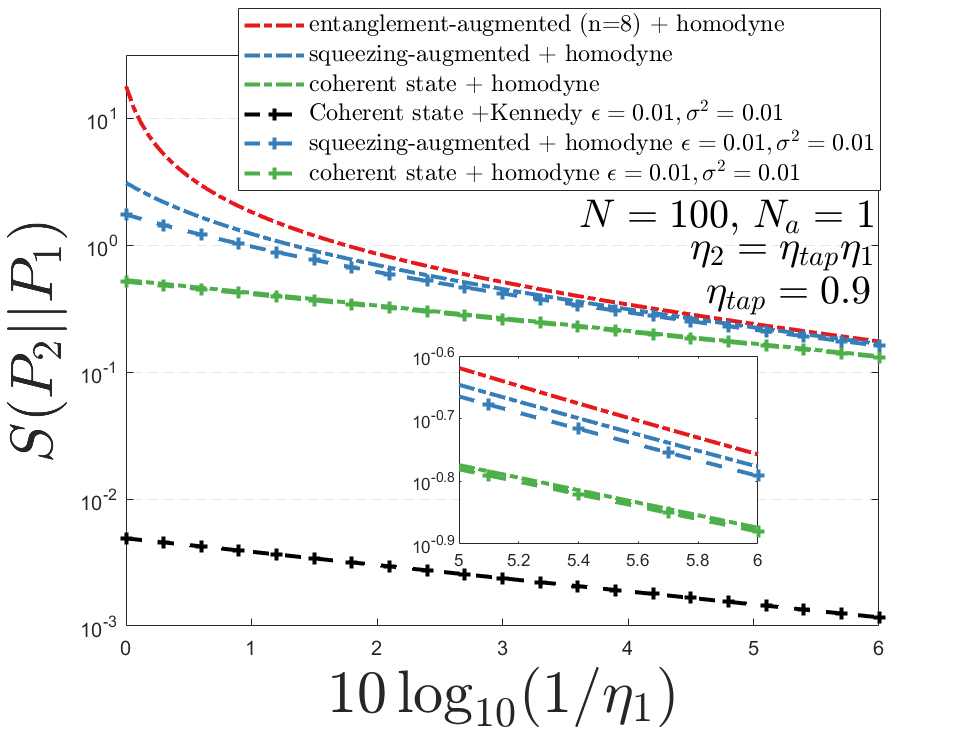}
\caption{
CRE as a function of the pre-change channel loss in dB. An ideal Kennedy receiver achieves CRE$=\infty$, but its CRE is very sensitive to amplitude and phase noise in the LO. An imperfect LO pulse is a coherent state $|\alpha_{\rm{LO}}(1+\epsilon)e^{i\theta}\rangle$, where $0 < \epsilon \ll 1$ is the {\em amplitude error} and $\theta \sim {\mathcal N}(0, \sigma^2)$, a Normal distributed random variable with variance $\sigma^2 \ll 1$ (mildly truncated to $[-\pi, \pi]$) is the {\em phase error}. Homodyne detection on the other hand is very robust to LO noise.}
\label{fig:KennedyReceiver}
\end{figure}

One might wonder, since there already exists a physically-realizable receiver that achieves infinite CRE with coherent state transmission (even with no quantum augmentation), why one should bother with the quantum augmentations described earlier in this Letter. The answer lies in the Kennedy receiver's CRE being enormously more sensitive---to tiny phase- and amplitude- errors in the local oscillator (LO)---compared to that of a Homodyne detection receiver. This extreme sensitivity to the nulling accuracy makes the infinite CRE practically unattainable. See Fig.~\ref{fig:KennedyReceiver} for numerical results, and Section~\ref{sec:endmatter}(a) for detailed analytical derivations of the LO-noise sensitivity of the CRE of both receivers.

{\em Conclusions and outlook}---We showed that spreading a tiny amount of continuous-variable entanglement across blocks of bright phase-modulated laser-light pulses, paired with a homodyne detection receiver at the output of a lossy channel, dramatically increases the sensitivity of detecting a sudden increase in the channel's loss, without affecting the (classical) communications performance. The quantum enhancement is the most pronounced when the initial channel loss is low, i.e., high $\eta_1$. All downstream losses after transmission can be subsumed into what we termed `initial loss'. The maximum reduction in the change-detection latency using our protocols is by a factor equaling the inverse of $1-\eta_1$. We showed that an ideal Kennedy receiver can in principle attain instantaneous change detection even with no quantum augmentation of the laser-light pulses, but that its performance is extremely sensitive to the local oscillator amplitude and phase noise, limiting its practical utility. We expect our results to extend to high-order modulation alphabets paired with heterodyne detection, and to noisy coherent states~\cite{Jagannathan2022}. In Appendix~\ref{app:sp_augmented}, we discuss single-photon augmentation of coherent state pulses. A more thorough future study of discrete-variable non-Gaussian quantum augmentations would be interesting. It is also of interest to consider the scenario when there are no (base) strong coherent state pulses on which the quantum-augmented photons would ride (i.e., $N=0, N_a > 0$). In other words, change detection of channel loss is the only relevant task with no underlying data communications. In this setting, we show that a pulsed single-photon source paired with a single photon detector outperforms a coherent-state source paired with homodyne detection (see Appendix~\ref{app:sp_augmented}). This naturally extends to a scenario of joint {\em quantum} communications (using dual-rail single-photon-qubit modulation) and change detection (see Appendix~\ref{sec:jointcommsensing}), which has been further studied in Ref.~\cite{Gong2025}. Finally, the problem of evaluating the optimum tradeoff of the (classical and/or quantum~\cite{Wilde2012}) communications rate and quantum-augmented quickest change detection of the channel's loss and other characteristics, and associated optimal transceiver-code-receiver combinations is left open as an avenue for future research.

\begin{acknowledgments}
This research was supported by the DARPA Quantum Augmented Networking (QuANET) Program under contract number HR001124C0405, and partially by the Office of Naval Research (ONR) under grant number N000142412627.
\end{acknowledgments}



\appendix

\section{Robustness Analysis of Homodyne and Kennedy receivers under Amplitude and Phase Imperfections in the Local Oscillator (LO)}~\label{sec:endmatter}

\begin{figure}[h]
\centering
\includegraphics[width=\columnwidth]{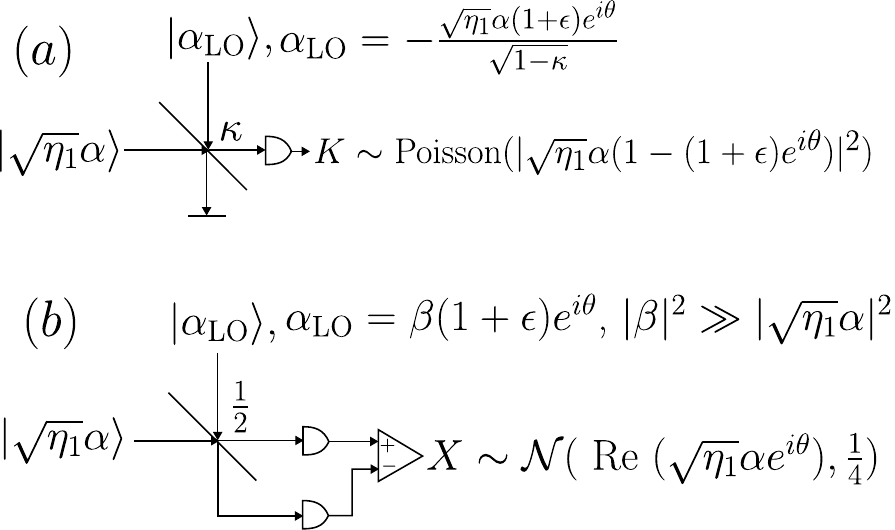}
\caption{$(a)$: Kennedy receiver. $(b)$: homodyne detection receiver. The Kennedy receiver uses a beam splitter with transmissivity $\kappa \approx 1$ and an LO of amplitude  $\alpha_{\rm LO} = -\frac{\sqrt{\eta_1}\alpha(1+\epsilon)e^{i\theta}}{\sqrt{1-\kappa}}$ to attempt nulling the received coherent state amplitude. The homodyne receiver uses a 50:50 beam splitter followed by two photodiodes, and an electronic-domain difference amplifier and integrator acting on the photocurrent outputs.}
\label{fig:receivers}
\end{figure}

Figure~\ref{fig:receivers} shows the constructions of the (a) Kennedy and (b) Homodyne receivers.
Both receivers use a coherent-state LO, with intended LO-amplitude denoted $\alpha_{\rm{LO}}$. Homodyne detection uses a strong LO, i.e., $|\alpha_{\rm{LO}}|^2 \gg |{\sqrt{\eta_1}\alpha}|^2$, and does {\em not} need to know the amplitude of the received pulse $\sqrt{\eta_1}\alpha$. The Kennedy receiver on the other hand mixes the received pulse $|\sqrt{\eta_1}\alpha\rangle$ in a beamsplitter of transmissivity $\kappa \approx 1$ with an LO pulse of amplitude $\alpha_{\rm{LO}} = \sqrt{\eta_1}\alpha/\sqrt{1-\kappa}$, intended to apply a phase-space displacement $D(-\sqrt{\eta_1}\alpha)$ to the received pulse, thus displacing it to vacuum, in the limit that $\kappa \to 1$. We model an imperfect LO, for either case above, as a coherent state $|\alpha_{\rm{LO}}(1+\epsilon)e^{i\theta}\rangle$, where $0 < \epsilon \ll 1$ is the {\em amplitude error} and $\theta \sim {\mathcal N}(0, \sigma^2)$, a Normal distributed random variable with variance $\sigma^2 \ll 1$ (mildly truncated to $[-\pi, \pi]$) is the {\em phase error}. 

The Kennedy receiver attempts to displace the pre-change received coherent state to the vacuum state. Let us begin with analyzing the effect of just an amplitude error in the Kennedy Receiver's local oscillator, $\epsilon$, to the coherent-state transmitter. The pre-change state, after the Kennedy Receiver's imperfect nulling-displacement becomes $ \ket{\psi_1}=\ket{-\epsilon}$ and the post-change state becomes $ \ket{\psi_2}=\ket{(\sqrt{\eta_2} -\sqrt{\eta_1})\alpha -\epsilon } $. These are detected using a photon-number-resolving (PNR) detector, whose output is an integer random variable $K$ that follows a Poisson distribution with means $\lambda_1 = \epsilon^2$ under $\ket{\psi_1} $ and $ \lambda_2 = ((\sqrt{\eta_2} -\sqrt{\eta_1})\alpha-\epsilon)^2 $ under $ \ket{\psi_2} $. The relative entropy $S(K_2||K_1)$ is given by:
\begin{align}
    &S(K_2||K_1)\nonumber\\
    &= \lambda_2\log\left(\frac{\lambda_2}{\lambda_1}\right) + \lambda_1-\lambda_2 \label{eq:RE} \\
    & = ((\sqrt{\eta_2} -\sqrt{\eta_1})\alpha-\epsilon)^2 \log\left( \frac{((\sqrt{\eta_2} -\sqrt{\eta_1})\alpha-\epsilon)^2}{\epsilon^2}\right) \nonumber\\
    &\quad+ 2\epsilon (\sqrt{\eta_2} -\sqrt{\eta_1})\alpha - ((\sqrt{\eta_2} -\sqrt{\eta_1})\alpha)^2 \label{eq:1}.
\end{align}
The first term in \eqref{eq:1} diverges as $\epsilon\to 0$. In particular, it is straightforward to show that $S(K_2||K_1)$ scales as $\ln(\epsilon)$ as $\epsilon\to 0$ in the limit $\epsilon \to 0$, i.e.,
\begin{align}
    \lim_{\epsilon\to 0}\frac{S(X_2||X_1)}{\ln(\epsilon)} = 2((\sqrt{\eta_2} -\sqrt{\eta_1})\alpha)^2.
\end{align}
Next, we consider the effect of phase error on the Kennedy Receiver's performance. The ideal intended LO amplitude $\alpha_{\rm LO}$ is replaced by $\alpha_{\rm LO}e^{i\theta}(1+\epsilon) $. The applied displacement therefore becomes $\hat{D}\left( -\sqrt{\eta_1}\alpha e^{i\theta}(1+\epsilon) \right)$. Let $p_{\lambda}(k) = e^{-\lambda} \frac{\lambda^k}{k!}$ denote the probability mass function of a Poisson distribution with mean $\lambda$, and $q_{\mu,\sigma^2}(x) = \frac{1}{\sqrt{2 \pi \sigma^2}}e^{-(x-\mu)^2/2\sigma^2}$ denote the probability density function of a Gaussian distribution $\mathcal{N}(\mu,\sigma^2)$. With amplitude and phase errors $\epsilon$ and $\theta$ respectively, the pre-change ($s=1$) and post-change ($s=2$) coherent states, when detected by PNR detection, yield a Gaussian-mixed Poisson distribution:  
\begin{align}
    p_s^{(\rm   Kennedy)}(k) = \int_{-\pi}^{\pi} q_{0,\sigma^2}(\theta) p_{\lambda_s(\theta)}(k) d\theta, \ s = 1,2, \label{eq:pdf kennedy}
\end{align}
where $\lambda_s(\theta) = |\alpha(\sqrt{\eta_s}- \sqrt{\eta_1}e^{i\theta}(1+\epsilon))|^2 $. The CRE plotted in Fig.~\ref{fig:KennedyReceiver} is evaluated numerically from \eqref{eq:pdf kennedy}.

\begin{figure}[tp]
\includegraphics[width=\columnwidth]{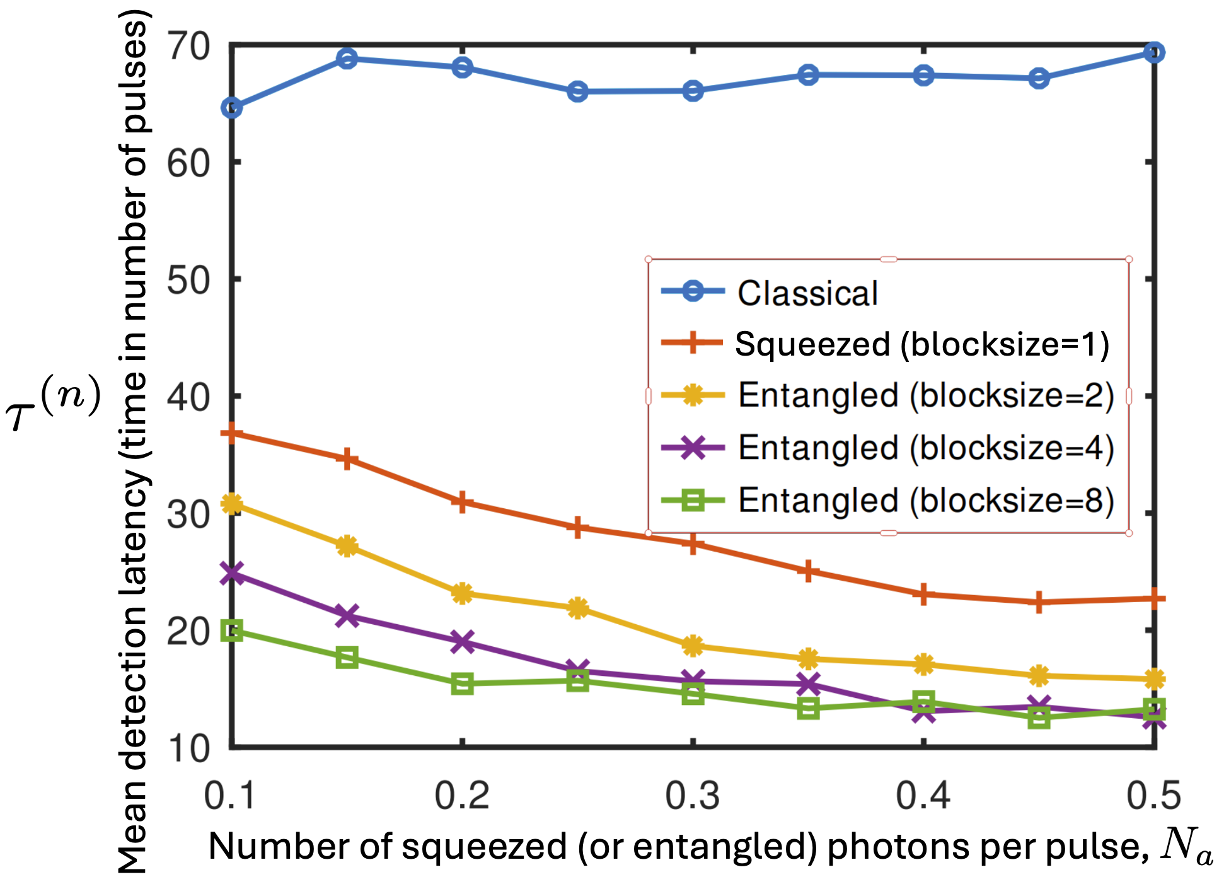}
\caption{Simulated mean change-detection latency $\tau^{(n)}$ plotted as a function of $N_a$, for the classical ($n=0$) and quantum-augmented ($n=1,2,4,8$) cases, generated using a CUSUM-variant algorithm that accounts for correlations within the $n$-blocks. The classical coherent mean photon number per pulse, $N = 100$. The change occurs at $t = 1005$, which is in the middle of an $n$-length entangled block, for each of $n=2, 4$ and $8$ considered here.}
\label{fig:ARL}
\end{figure}

For homodyne detection, we will evaluate the effect of LO imperfections for the squeezing-augmented transmitter $(n=1)$, which subsumes the coherent-state transmitter ($r=0$) as a special case. The amplitude error $\epsilon$ has no effect since $|\alpha_{\rm LO}|^2 \gg |\sqrt{\eta_1}\alpha|^2$.
The LO phase error $\theta$ corresponds to a random phase rotation applied to the received coherent state. The resulting distribution of $X$, the homodyne receiver output, is given by:
\begin{align}
    p_{s}^{(\rm Homodyne)}(x) = \int_{-\pi}^{\pi} q_{0,\sigma^2}(\theta) q_{\mu(\theta),\Sigma_s(\theta)}(x) d\theta, \ s = 1,2, \label{eq:pdf homodyne}
\end{align}
where $ 
    \mu_s =      \cos \theta \sqrt{\eta_s}\alpha $,  $
    \Sigma_s  =\Sigma_{s-}\cos^2\theta+\Sigma_{s+}\sin^2\theta $, with $
    \Sigma_{s+} = \frac{1}{4}((1+2\Bar{n}_{\rm B})(1-\eta_s)+\eta_se^{2r}) $, $ 
    \Sigma_{s-} = \frac{1}{4}((1+2\Bar{n}_{\rm B})(1-\eta_s)+\eta_se^{-2r}) $. The CREs $S(X_2||X_1)$ shown in Fig.~\ref{fig:KennedyReceiver} are evaluated numerically from \eqref{eq:pdf homodyne}. We omit the numerical analysis of the entanglement-augmented transmitters. In Fig.~\ref{fig:ARL} we plot detection latencies $\tau^{(n)}$, for $n=0$ (coherent-state) and $n=1, 2, 4$ and $8$ (quantum-augmented) transmitters, obtained via Monte Carlo CUSUM simulations, as a function of $N_a$. The ratios $\tau^{(n)}/\tau^{(n^\prime)}$ are consistent with the respective theoretical CRE ratios $S^{(n^\prime)}/S^{(n)}$, $\forall n,n^\prime$. For the simulations, we update the CUSUM counter using the conditional log-likelihood ratios, rather than the conventional log-likelihood ratios employed in standard CUSUM tests. We describe our algorithm in further detail in Appendix~\ref{sec:blockcusum}.

\section{Squeezing-augmented transmitter}\label{app:sq_augmented}

\subsection{Generation of squeezing-augmented BPSK-modulated codewords}\label{app:squeezingaugmented_transmitter_schematic}

\begin{figure}[h]
\centering
\includegraphics[width=\columnwidth]{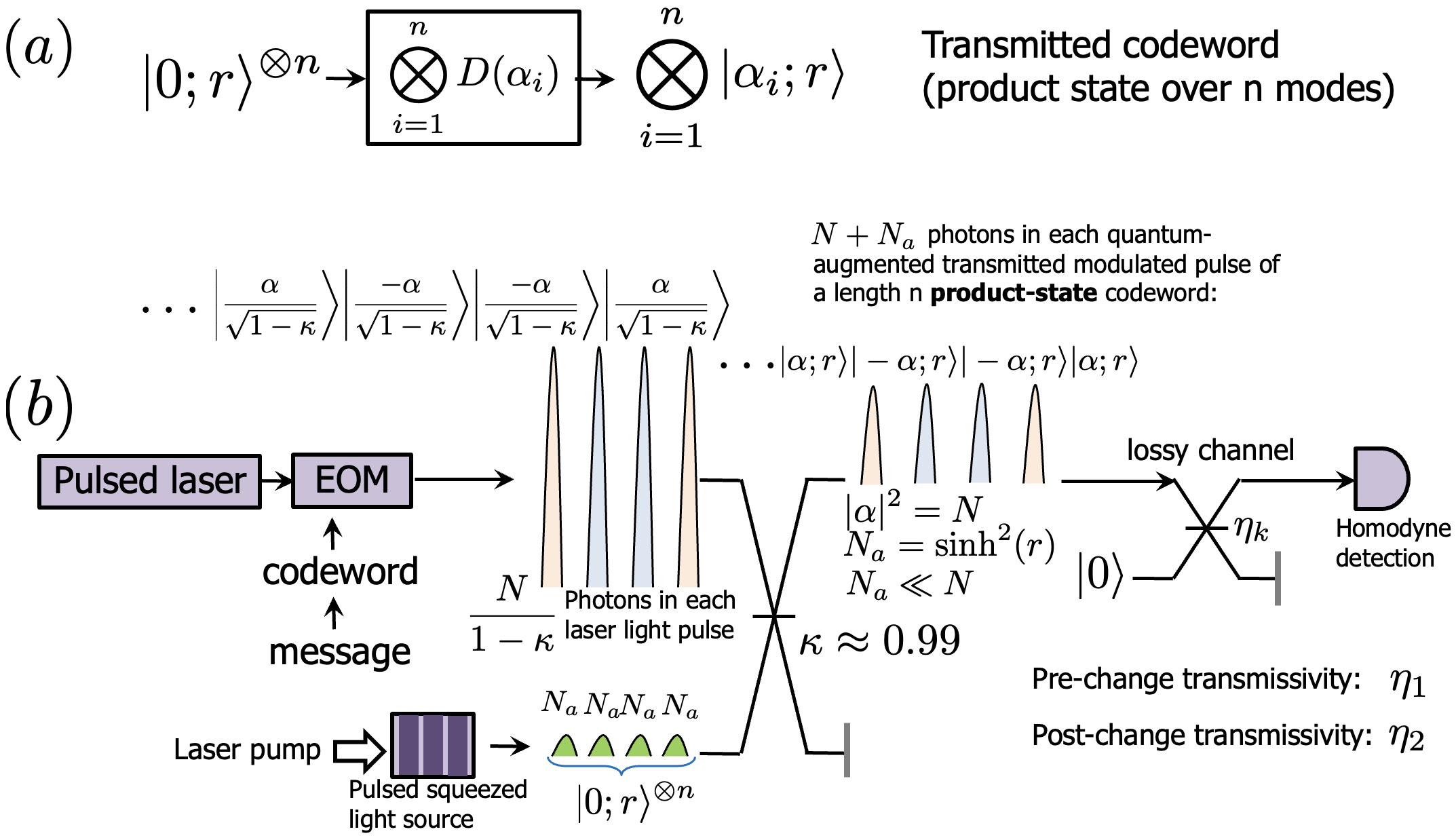}
\caption{Transmitter generating squeezing-augmented coded-BPSK-modulated laser-light pulses. Also shown is homodyne detection at receiver, at the output of a lossy channel.}
\label{fig:System_SqueezingAugmentation}
\end{figure}

Fig.~\ref{fig:System_SqueezingAugmentation} shows a schematic of one way of how our proposed squeezing-augmented BPSK coherent communications transceiver system could be built. We encode the message to be transmitted over the channel using a binary code and generate the corresponding modulated coherent-state BPSK codewords. Each pulse has a (large) mean photon number $N/(1-\kappa)$, for $\kappa$ taken to be very close to $1$ (such as $0.99$). We mix pulses of this train of strong-energy coded-modulated BPSK pulses with squeezed-vacuum pulses each of mean photon number $N_a$, on a $\kappa$-transmissivity beam splitter. This results in generating our desired train of coded-modulated squeezing-augmented BPSK pulses (displaced squeezed states $|\alpha; r\rangle$ or $|-\alpha; r\rangle$), each with mean photon number $\approx N+N_a$, which is then sent over the lossy communication channel. Another way to generate this sequence of binary-phase-modulated displaced squeezed-state pulses would be to seed the input facet of the nonlinear crystal used to generate the squeezed-vacuum pulses (e.g., via spontaneous parametric downconversion) with BPSK modulated laser-light pulses at the squeezed frequency, which would result in generating BPSK-modulated displaced squeezed pulses (rather than squeezed-vacuum pulses), directly. The homodyne detection receiver at the output serves two purposes---decoding the transmitted message, and detection of a sudden change in the channel's transmissivity (quicker than it would be possible without the quantum augmentation).

\subsection{Derivative of $S^{(1)}$ at $N_a = 0$}~\label{sec:derivative}

The derivative $\partial S^{(1)}/\partial N_a$ is given by $(\partial S^{(1)}/\partial r)/ (\partial N_a/\partial r)$, where $N_a = {\rm sinh}^2(r)$. Using $\partial N_a/\partial r = 2{\rm sinh}(r){\rm cosh}(r)$, and the expression of $S^{(1)}$ from Eq.(3) from the Letter, it is simple to evaluate the following:
\begin{equation}
\frac{\partial S^{(1)}}{\partial N_a} = \frac{e^{-2r}(2\eta_1N(\sqrt{\eta_2}-\sqrt{\eta_1})^2 + 4v_2\eta_1 - 4v_1\eta_2)}{(4v_1)^2{\rm sinh}(r){\rm cosh}(r)},
\end{equation}
where $v_j = [\eta_j e^{-2r} + (1-\eta_j)]/4$, $j = 1$ or $2$. At $N_a = 0$ (i.e., $r=0$), $v_1 = v_2 = 1/4$, resulting in $({\partial S^{(1)}}/{\partial N_a})|_{N_a=0} = (2\eta_1N(\sqrt{\eta_2}-\sqrt{\eta_1})^2 + \eta_1 - \eta_2)/{\rm sinh}(r)|_{r=0}$, which is $\infty$ at $r=0$. This is in sharp contrast with $\partial S^{(0)}/\partial N_a = 2({\sqrt{\eta_2}}-{\sqrt{\eta_1}})^2$ for the classical case, and explains the sharp increase in the relative entropy (and hence the sharp decrease in the change-detection latency) when a tiny amount of squeezing-augmentation is used. Interestingly, this infinite derivative of $S^{(1)}$ with respect to $N_a$ is independent of $N$, and hence would work even for bright laser-light modulation deployed in standard optical communications networks.

\subsection{Threshold $N_a$ above which squeezing augmentation does not outperform classical baseline}

There is diminishing return to adding squeezing for augmenting bright coherent state pulses for loss-change detection. Since $S^{(0)}$ increases linearly with $N_a$ and $S^{(1)}$ saturates at $2N(\sqrt{\eta_2}-\sqrt{\eta_1})^2/(1-\eta_1)$, when $N_a > N_{a,{\rm th}}$, $S^{(1)} < S^{(0)}$, i.e., squeezing augmentation does not help for comparable-energy classical transmission, where as we will show in this section, the threshold $N_{a,{\rm th}} \approx N\eta_1/(1-\eta_1)$.

When $N$ is large, the second term in the expression of $S^{(1)}$ in Eq.(3) from the Letter proportional to $N$ dominates; so we will ignore the first term in Eq.(3) from the Letter for this phenomenological derivation of the threshold. Further, we are interested in the high-$N_a$, therefore high-$r$ regime for the purposes of this appendix. Therefore, we will assume:
\begin{equation}
S^{(1)} \approx \frac{2N(\sqrt{\eta_2} - \sqrt{\eta_1})^2}{1 - \eta_1},
\end{equation}
whereas
\begin{equation}
S^{(0)} = 2(N+N_a)(\sqrt{\eta_2} - \sqrt{\eta_1})^2.
\end{equation}
squeezing augmentation does not help for comparable-energy classical transmission for $N_a > N_{a,{\rm th}}$, where $N_{a,{\rm th}} \approx N\eta_1/(1-\eta_1)$. Since we are assuming $N$ is large for the purposes of this derivation, and since we know that the high-$\eta_1$ regime is where squeezing augmentation would provide the most value, $N_{a,{\rm th}}$ would typically be too high anyway for it to correspond to reasonable single-mode squeezing. For example, for $N=100$ and $\eta_1 = 0.9$, $N_{a,{\rm th}} = 900$, which corresponds to roughly $35$ dB of squeezing, which is well beyond anything that can be practically generated. This ascertains that the regime where squeezing augmentation helps is the small-signal quantum augmentation regime, i.e., $N_a \ll N$.

\subsection{Block CUSUM test}~\label{sec:blockcusum}

For joint communication and change detection using an entanglement-augmented coherent state,the message is decoded after the entire block of the message has been received. In this block-wise scheme, the log-likelihood ratio is accumulated over the block, and a change can be declared only at the block boundaries. In contrast, our approach updates the conditional log-likelihood ratio immediately after each message symbol is received. This strategy is not provably optimal in the same sense as the standard CUSUM test.

Consider an i.i.d.~sequence of random variable vectors $Y_1^{(s)}, Y_2^{(s)},\dots$, where $s=1,2$ denotes prechange and postchange. The random variable vector $Y^{(s)} = (X_{1}^{(s)},\dots,X_{m}^{(s)})$. We consider the case where the components of the m-vector arrive sequentially in time. That is, $X_{1:j-1}^{(s)} = \{X_{1}^{(s)}, X_{2}^{(s)},\dots,X_{j-1}^{(s)}\}$ are observed prior to $X_{j}$. The $m\times1$ mean vector and the $m\times m$ covariance matrix of $Y^{(s)}$ are given by
\begin{align}
    \bm{\mu^{(s)}} & = \begin{bmatrix}
&\bm{\mu^{(s)}}_{1:j-1} \\
& \mu^{(s)}_{j} \\
& \bm{\mu^{(s)}}_{j+1:m}
    \end{bmatrix}, \
    \Sigma_s &= \begin{bmatrix}
    &\Sigma^{(s)}_{1:j-1,1:j-1} & \Sigma^{(s)}_{1:j-1,j} & \cdots\\
&\Sigma^{(s)}_{j,1:j-1} & \Sigma^{(s)}_{j:j} & \cdots\\
& \vdots & \vdots & \ddots
\end{bmatrix},\nonumber
\end{align}
where $\bm{\mu^{(s)}}_{k:l} = [\mu^{(s)}_{k},\dots,\mu^{(s)}_{l}]^T $ is a $ l-k+1$ by $1$ vector and \begin{align}
    \Sigma^{(s)}_{k_1:l_1,k_2:l_2}  = \begin{bmatrix}
        \Sigma^{(s)}_{k_1,k_2} & \dots & \Sigma^{(s)}_{k_1,l_2} \\
         \vdots & \ddots & \vdots \\
        \Sigma^{(s)}_{l_1,k_2} & \dots & \Sigma^{(s)}_{l_1,l_2} \\
    \end{bmatrix}
\end{align} is $ l_1-k_1+1 $ by $ l_2-k_2+1 $ matrix.
The conditional random variable $X_j^{(s)}|X_{1:j-1}^{(s)}\sim\mathcal{N}\left(\mu^{(s)}_{j|1:j-1}, \sigma^{2(s)}_{j|1:j-1} \right)$ is Gaussian distributed, where
\begin{align}
\mu^{(s)}_{j|1:j-1} & = \mu^{(s)}_{j}+ \Sigma^{(s)}_{j,1:j-1} \Sigma_{-}^{(s)} \left( \bm{x}_{1:j-1} - \bm{\mu^{(s)}}_{1:j-1} \right) \\
\sigma^{2(s)}_{j|1:j-1} &= \Sigma^{(s)}_{j,j} - \Sigma^{(s)}_{j,1:j-1}\Sigma_{-}^{(s)}\Sigma^{(s)}_{1:j-1,j} \\
\Sigma_{-}^{(s)}&=\left( \Sigma^{(s)}_{1:j-1,1:j-1} \right)^{-1}.
\end{align}
The CUSUM counter at time $j$ is $L_j = \max\left( 0, L_{j-1}+\log\frac{p^{(2)}}{p^{(1)}} \right)$, where $p^{(s)}$ is the conditional probability distribution function of $ X_j^{(s)}|X_{1:j-1}^{(s)}\sim\mathcal{N}\left(\mu^{(s)}_{j|1:j-1}, \sigma^{2(s)}_{j|1:j-1} \right) $.

\section{Continuous-variable entanglement augmented transmitter}~\label{sec:CVtransmitter}

\subsection{Generation of entanglement-augmented BPSK-modulated codewords}\label{app:entangled_transmitter_schematic}
\begin{figure}[h]
\centering
\includegraphics[width=\columnwidth]{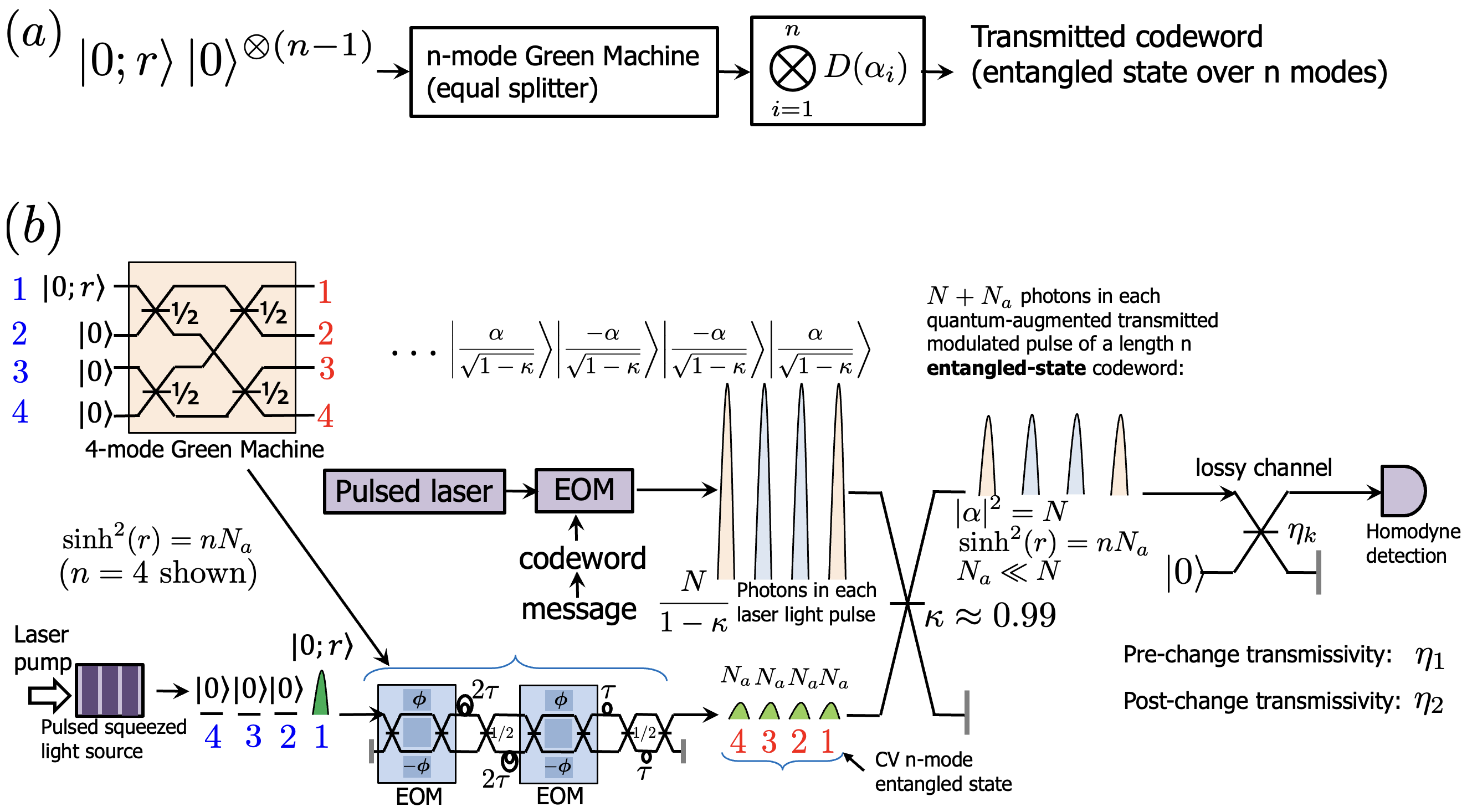}
\caption{Transmitter generating entanglement-augmented coded-BPSK-modulated laser-light pulses. Also shown is homodyne detection at receiver, at the output of a lossy channel.}
\label{fig:System_EntanglementAugmentation}
\end{figure}
Fig.~\ref{fig:System_EntanglementAugmentation} shows a schematic of how our proposed entanglement-augmented communications transceiver system could be built. We generate blocks of $n$-mode continuous-variable entangled states over $n$ pulse slots by passing a train of squeezed vacuum pulses $|0;s\rangle$ (with ${\rm sinh}^2(s) = nN_a$)---generated at an $n$-fold-lower repetition rate compared to the intended repetition rate of the modulated communications pulses---through the temporal-mode Green Machine~\cite{Cui2023}. In parallel, we generate the modulated coherent-state BPSK codewords, for the transmitted raw bit sequence and our choice of the error correction code, but with each pulse having a (large) mean photon number $N/(1-\kappa)$, for $\kappa$ taken to be very close to $1$ (such as $0.99$). We then mix consecutive blocks of $n$ pulses of this train of strong-energy coded-modulated BPSK pulses with blocks of $n$ entangled pulses at the output of the Green Machine, on a $\kappa$-transmissivity beam splitter. This results in generating our desired train of coded-modulated entanglement-augmented BPSK pulses, each with mean photon number $\approx N+N_a$, which is then sent over the lossy communication channel. The homodyne detection receiver at the output serves two purposes---decoding the transmitted message, and detection of a sudden change in the channel's transmissivity (quicker than it would be possible without the quantum augmentation).

\subsection{Derivation of the statistics of the homodyne-receiver output}
\label{app:ent_augmented}
We describe the $n$-mode case of the entanglement-augmented transmitter in this section. We will assume augmenting unmodulated coherent states for simplicity of exposition. We will consider the $n$-mode Hadamard unitary for the derivation, also known as the {\em Green Machine}~\cite{Guha2011}. Since the Green machine acts on $2^k$-modes, for $k \in {\mathbb Z}$ we  take $n=2^k$ for the analysis in this Appendix. However, we could equally well use the complex-valued-Hadamard linear-optical unitary, which exists for all $n$, or for that matter any equal-power linear-optical $n$-splitter. 

\begin{figure}[h]
\begin{tikzpicture}
\draw[->] (-5.1,0)-- (-4,0) node[midway, above] {$\ket{0;r}$};
\draw[->] (-5.1,-1.5)-- (-4,-1.5) node[midway, above] {$\ket{0}$};
\node at (-4.6,-2.3) {\vdots};
\node at (-1.5,-2.3) {\vdots};
\node at (0.0,-2.3) {\vdots};
\draw[->] (-5.1,-3.8)-- (-4,-3.8) node[midway, above] {$\ket{0}$};
 \draw[thick] (-4,-4.3) rectangle (-3,0.5) node at (-3.5,-2) {$H_{2^k}$};
 \draw[->] (-3,0)-- (-2,0);
 \draw[->] (-3,-1.5)-- (-2,-1.5);
  \draw[->] (-3,-3.8)-- (-2,-3.8);
 \draw[thick] (-2,-0.5) rectangle (-1,0.5) node at (-1.5,0) {$D(\alpha)$};
  \draw[thick] (-2,-2) rectangle (-1,-1) node at (-1.5,-1.5) {$D(\alpha)$};
    \draw[thick] (-2,-4.3) rectangle (-1,-3.3) node at (-1.5,-3.7) {$D(\alpha)$};
\draw[->] (-1,0)-- (-0.1,0) node[midway, above] {};
\draw[->] (0.1,0) -- (1.1,0) node[near end,above] {}; 
\draw[->] (0,1) -- (0,0.1) node[midway, right] {$\ket{0}$};
\draw[-] [thick] (-0.5,0.5) -- (0.5,-0.5); 
\node at (0.5, -0.3) {$\eta$};
\draw[-] [thick] (-0.5,-3.3) -- (0.5,-4.3); 
\node at (0.5, -4.1) {$\eta$};
\draw[->] (-1,-1.5)-- (-0.1,-1.5) node[midway, above] {};
\draw[->] (-1,-3.8)-- (-0.1,-3.8) node[midway, above] {};
\draw[->] (0.1,-1.5) -- (1.1,-1.5) node[near end,above] {}; 
\draw[->] (0.1,-3.8) -- (1.1,-3.8) node[near end,above] {}; 
\draw[->] (0,-0.5) -- (0,-1.4) node[midway, right] {$\ket{0}$};
\draw[->] (0,-2.9) -- (0,-3.7) node[midway, right] {$\ket{0}$};
\draw[-] [thick] (-0.5,-1) -- (0.5,-2); 
\node at (0.5, -1.8) {$\eta$};
\draw [thick] [decorate, decoration = {calligraphic brace}] (1.3,0.1) --  (1.3, -3.9);
\node at (2.5,-1.4){Real};
\node at (2.5,-1.8){quadrature};  
 \node at (2.5,-2.2){Homodyne};
 \node at (2.5,-2.5){detection};
\end{tikzpicture}
 \caption{$n=2^k$-mode entanglement-augmented transmitter }
 \label{fig:entangled}
 \end{figure}
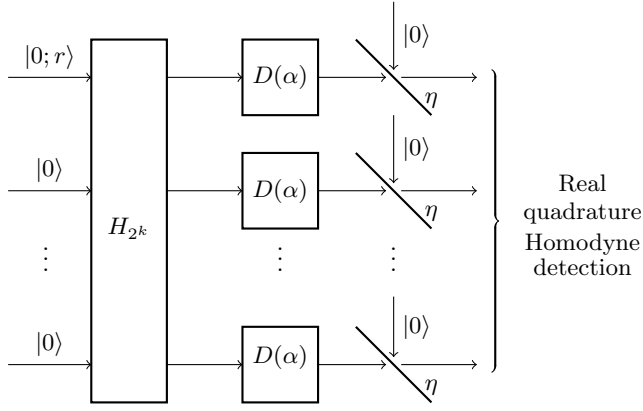

Figure \ref{fig:entangled}
shows our setting.
Now we set out to prove  Equations (4)-(6) in the Letter. Let $\hat{\rho}$ be the state at the output of the lossy channel with transmissivity $\eta$.    Note that  the mean of the real quadrature Homodyne detection is the real part of the mean of the state  being measured. Also, the second moments of the real quadrature Homodyne detection outputs are exactly the same as the $q_lq_l$ and $q_lq_m$ covariances of the state being measured. In other words, if $X_l$ denotes the random variable of the $l$-th real quadrature Homodyne detection (i.e., measurement of the $l$-th momentum operator $\hat{q}_l$), $l=1,2,\dots,n$, then the mean and covariances of $X_l$ and $X_m$ are given by  \begin{align}\begin{split}
E[X_l]&=\Re\left( \text{Tr}\left(\hat{\rho}\hat{q}_l\right)\right),\\
E[\Delta_l^2]&=\text{Tr}\left(\hat{\rho}\left(\hat{q}_l-\text{Tr}\left(\hat{\rho}\hat{q}_l\right)\right)^2\right),\quad \text{ and}\\
E[\Delta_l\Delta_m]&=\text{Tr}\left(\hat{\rho}\left(\hat{q}_l-\text{Tr}\left(\hat{\rho}\hat{q}_l\right)\right)\left(\hat{q}_m-\text{Tr}\left(\hat{\rho}\hat{q}_m\right)\right)\right),\end{split}\label{eq:appendix-mean-cov}
\end{align}where $\Delta_l = X_l-E[X_l]$, $l=1,2,\dots,n$. Since $\alpha$ is a real number to prove Equations (4), (5) and (6) from the Letter, 
it is enough to show that the state at the output of the pure loss channel with transmissivity $\eta$ has; \begin{enumerate}
    \item \label{item:mean} mean given by $\sqrt{\eta}\alpha$ in each mode; and 
    \item \label{item:2-moments} the covariance matrix has its entries given by $\frac{e^{-2r}+n-1}{4n}$ at the $\hat{q}_l\hat{q}_l$-positions,  and  $\frac{e^{-2r}-1}{4n}$ at $\hat{q}_l\hat{q}_m$-positions, $j,k=1,\dots,n$ and $j\neq k$.
\end{enumerate}
The proof of \ref{item:mean} is easy because the mean of the state at the input of the pure loss channel is $\alpha$ at each mode.
 Now we prove \ref{item:2-moments} using mathematical induction on $k$.
Let $V_{1}= \frac{1}{4}\begin{bmatrix}
e^{-2r}&0\\0&e^{2r}
\end{bmatrix}$, the covariance matrix of the squeezed vacuum state $\ket{0,r}$, with the convention $\hbar=\frac{1}{2}$. Now for $k>0$, let \[V_{2^k}=V_{2^{k-1}}\oplus \frac{1}{4} I_{2^k},\] a $2^{k+1}\times 2^{k+1}$ matrix, which is the covariance matrix of the state $\ket{0,r}\otimes \ket{0}^{\otimes^{2^k-1}}$. Now let $H_1=\begin{bmatrix}
    1&0\\0&1
\end{bmatrix}$ and for $k>0$, let \[H_{2^k} =\frac{1}{\sqrt{2}}\begin{bmatrix}
    H_{2^{k-1}}&H_{2^{k-1}}\\H_{2^{k-1}}&-H_{2^{k-1}}
\end{bmatrix}, \] a $2^{k+1}\times 2^{k+1}$ orthogonal matrix, which is the real form of the $k$-th Hadamard unitary matrix. We claim that 
\onecolumngrid
\begin{align}\label{eq:out-at-green}
 V&:= H_{2^k}V_{2^k} H_{2^k}^T\nonumber\\&= \scalebox{1}{$\frac{1}{2^{k+2}}
\begin{bmatrix} 
e^{-2r}+2^{k}-1&0&e^{-2r} - 1&0&e^{-2r} - 1&\dots&e^{-2r}-1&0\\  0&e^{2r}+2^{k}-1&0&e^{2r} - 1&0&\dots&0&e^{2r}-1 \\e^{-2r}-1&0&e^{-2r}+2^{k}-1&0&e^{-2r}-1&\dots&e^{-2r}-1&0\\ \vdots&\vdots&\vdots&\vdots&\vdots&\vdots&\vdots&\vdots\\
 0&e^{2r}-1&0&e^{2r}-1&0&\dots&0&e^{2r}+2^{k}-1
\end{bmatrix}.$}
\end{align}
\twocolumngrid
We prove \ref{eq:out-at-green} using induction. To use induction, note first that, when $k=0$, \[H_1V_1H_1^T=V_1,\] satisfies the claim. Now assume that the claim is true for $k-1$, for some $k>0$, now we will prove that the claim is true for $k$. We have, by definition, \begin{align*}
   & H_{2^k}V_{2^k} H_{2^k}^T\\&=\frac{1}{2}\begin{bmatrix}
  H_{2^{k-1}}V_{2^{k-1}}H_{2^{k-1}}^T+\frac{1}{4}I_{2^k}&H_{2^{k-1}}V_{2^{k-1}}H_{2^{k-1}}^T-\frac{1}{4}I_{2^k}\\H_{2^{k-1}}V_{2^{k-1}}H_{2^{k-1}}^T-\frac{1}{4}I_{2^k}&H_{2^{k-1}}V_{2^{k-1}}H_{2^{k-1}}^T+\frac{1}{4}I_{2^k}  
\end{bmatrix}.
\end{align*}
By the induction hypothesis, the diagonal entries of the matrix $H_{2^{k-1}}V_{2^{k-1}}H_{2^{k-1}}^T$ are $\frac{1}{2^{k+1}}\left(e^{-2r}+2^{k-1}-1\right)$ and $\frac{1}{2^{k+1}}\left(e^{2r}+2^{k-1}-1\right)$ at $q_jq_j$ (odd) and $p_jp_j$ (even) positions respectively, and the non-diagonal entries are $\frac{1}{2^{k+1}}\left(e^{-2r} - 1\right)$,  $\frac{1}{2^{k+1}}\left(e^{2r} - 1\right)$ and $0$, respectively, for $q_jq_k,$ $p_jp_k$, and $q_jp_k$ (and $p_kq_j$) entries with $j\neq k$. Now, the claim is proved for $k$ by noticing that (i) the nondiagonal entries of $ H_{2^{k-1}}V_{2^{k-1}}H_{2^{k-1}}^T\pm\frac{1}{4}I_{2^k}$ are the same as those of $H_{2^{k-1}}V_{2^{k-1}}H_{2^{k-1}}^T$; and (ii) the equalities \begin{align*}
    \frac{1}{2}\left[\frac{e^{\pm2r}+2^{k-1}-1}{2^{k+1}}+\frac{1}{4}\right]&=\frac{e^{\pm2r}+2^{k}-1}{2^{k+2}}\\&=\frac{e^{\pm2r}+n-1}{4n}\\ \frac{1}{2}\left[\frac{e^{\pm2r}+2^{k-1}-1}{2^{k+1}}-\frac{1}{4}\right]&=\frac{e^{\pm2r}-1}{2^{k+2}}\\
    &=\frac{e^{\pm2r}-1}{4n}
\end{align*} hold.

If a state with mean vector $\bm{\alpha}=(\alpha_1,\dots,\alpha_{2^{k}})$ and covariance matrix given by \eqref{eq:out-at-green}  is transmitted through a lossy channel with a vacuum environment and transmissivity $\eta$, then the mean vector of the state  at the output is $\sqrt{\eta}\bm{\alpha}$ and the covariance matrix at the output is  \[\frac{1-\eta}{4}+\eta V,\]
where $V$ is as in Equation \eqref{eq:out-at-green}. As we discussed in Equation \eqref{eq:appendix-mean-cov} this proves Eqs.(4)-(6) from the Letter.

\section{Discrete-variable single-photon quantum augmentation}
\label{app:sp_augmented}
In this section, we will consider augmenting each (bright) coherent state pulse with a one-photon Fock state (i.e., $N_a=1$), by displacing the Fock state $|1\rangle$ by $D(\alpha)$. For the BPSK modulated transmitter, displacements of $D(\alpha)$ or $D(-\alpha)$ would be applied to a train of single photon pulses. First, let us calculate the output probability distribution of the homodyne detection receiver after lossy transmission of the state $D(\alpha)|1\rangle$ through a pure-loss channel of transmissivity $\eta$.
 \begin{center}
\begin{tikzpicture}
 \draw[->] (-3,0)-- (-2,0) node[midway, above] {$\ket{1}$};
 \draw[thick] (-2,-0.5) rectangle (-1,0.5) node at (-1.5,0) {$D(\alpha)$};
\draw[->] (-1,0)-- (-0.1,0) node[midway, above] {};
\draw[->] (0.1,0) -- (1.1,0) node[near end,above] {}; 
\draw[->] (0,1) -- (0,0.1) node[midway, right] {$\ket{0}$};
\draw[-] [thick] (-0.5,0.5) -- (0.5,-0.5); 
\node at (0.5, -0.3) {$\eta$};
\end{tikzpicture}
\end{center}
  Let $\ket{\psi}$ denote the state at the output of the pure loss channel with input state $D(\alpha)\ket{1}$, $\alpha\in \mathbb{R}$. Then the state $\hat{\rho}$ at the output is given by 
$\hat{\rho}=\text{Tr}_2 \ket{\psi}\bra{\psi}$,
where \begin{align}
\ket{\psi}:=\left[\Gamma(U_{\eta})D\left(\alpha\right)\otimes D(0)\right]\ket{1,0}.
\end{align}
The Fourier transform \footnote{Note that  in the convention used in this paper the Fourier transform of a distribution $\mu$ on the real line is given by integrating the function $e^{-itx}$ with respect to $\mu(dx)$} of the outcome distribution of the real quadrature Homodyne detection is given by: \begin{align*}
    \chi(t) &= \text{Tr}\left[ \left(\text{Tr}_2 \ket{\psi}\bra{\psi}\right) D\left(\frac{-it}{2}\right)\right]\\
    &= \bra{\psi}D\left(\frac{-it}{2}\right)\otimes D(0)\ket{\psi}\\
    &= \bra{1,0}D\left(\begin{bmatrix}
        \alpha\\0
\end{bmatrix}\right)^{\dagger}\Gamma(U_{\eta})^{\dagger}D\left(\begin{bmatrix}
        \frac{-it}{2}\\0
\end{bmatrix}\right) \nonumber \\
&\phantom{........}\times\Gamma(U_{\eta})D\left(\begin{bmatrix}
        \alpha\\0
\end{bmatrix}\right)\ket{1,0}.
\end{align*}
Since $\Gamma(U_{\eta})^{\dagger}D\left(\begin{bmatrix}
        \frac{-it}{2}\\0
\end{bmatrix}\right)\Gamma(U_{\eta}) = D\left(\begin{bmatrix}\frac{-it\sqrt{\eta}}{2}\\\frac{-it\sqrt{1-\eta}}{2}\end{bmatrix}\right)$, we now have: \begin{align*}
    \chi(t)&= \exp\{-it\sqrt{\eta}\alpha \}\bra{1}D\left(\frac{-it\sqrt{\eta}}{2}\right)\ket{1} \nonumber \\
    &\phantom{........}\times\bra{0}D\left(\frac{-it\sqrt{1-\eta}}{2}\right)\ket{0}\\
    &=\exp\{-it\sqrt{\eta}\alpha \}\exp\left\{\frac{-t^2\eta}{8} \right\}\nonumber \\
    &\phantom{........}\times\left(1-\frac{t^2\eta}{4}\right)\exp\left\{\frac{-t^2(1-\eta)}{8} \right\}\\
    &=\left(1-\frac{t^2\eta}{4}\right)\exp\left\{-it\sqrt{\eta\alpha} -\frac{t^2}{8}\right\}.
\end{align*}
Hence, the probability density function $p(t)$ of the output distribution is obtained by applying the inverse Fourier transform to $\chi(t)$. Thus, we have: \begin{align}\label{eq:pdfDV}
    p(x)&:= \sqrt{\frac{2}{\pi }} e^{-2 \left(x-\alpha  \sqrt{\eta }\right)^2} \left(4 \alpha ^2 \eta ^2+\eta  \left(4 x^2-1\right) \right. \nonumber \\
    &\phantom{........} \left. -8 \alpha  \eta ^{3/2} x+1\right),
\end{align}
for all $x\in \mathbb{R}.$

\begin{figure}[t]
\centering
\includegraphics[width=\columnwidth]{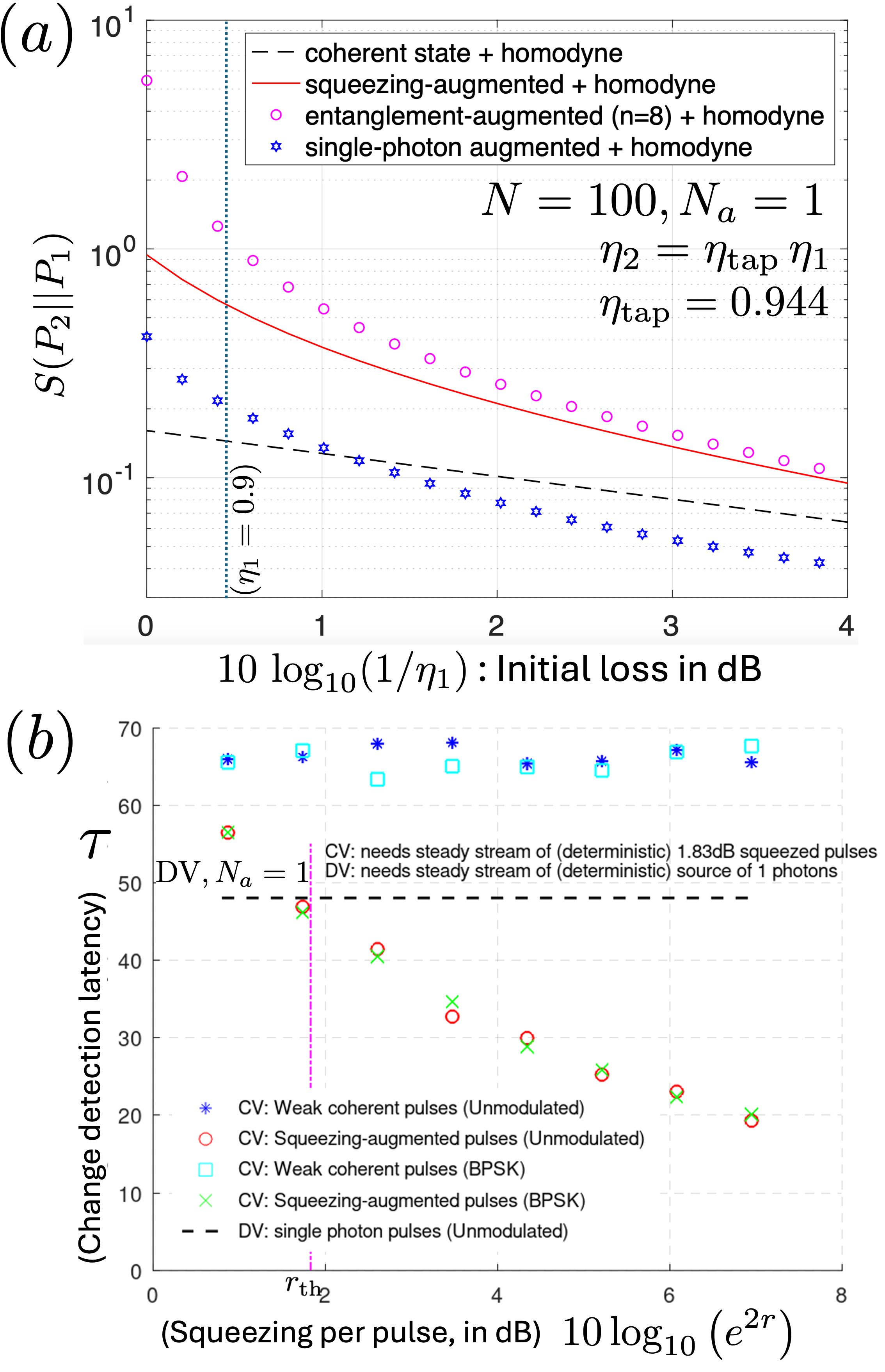}
\caption{(a) CRE versus pre-change loss in dB for the coherent state transmitter without quantum augmentation (black dashed), DV single-photon augmentation (blue hexagon markers), CV squeezed-light augmentation (red solid plot) and $n=8$ CV entanglement-augmentation (magenta circles). (b) Simulated values of the Change detection latency $\tau$ plotted as a function of the varying squeezing level (for CV) and the latency for the DV augmentation (single photon pulses each displaced by $D(\alpha)$, i.e., $N = \alpha^2$ and $N_a=1$. Each point is a mean over $500$ Monte Carlo runs. Simulations of CV vs. DV latency performance agree very well with the crossover at the theoretically predicted squeezing threshold $r_{\rm th}$ corresponding to roughly $1.83$ dB squeezing per pulse, as discussed in the text.}
\label{fig:DVaugmentation_sims}
\end{figure}

Using this distribution, we numerically calculate the CRE $S(P_2||P_1)$ for the discrete-variable (DV) augmentation scheme described above, and plot the CRE as a function of the pre-change loss in dB, in Fig.~\ref{fig:DVaugmentation_sims}(a). All the plots assume $N=100$ and $N_a=1$. The receiver is homodyne detection for all four plots. We assume $\eta_{\rm tap} = 0.944$, such that for $\eta_1 = 0.9$, we have $\eta_2 = 0.85$---the values assumed for the simulated detection latencies plotted in Fig.~\ref{fig:DVaugmentation_sims}(b). At $\eta_1 = 0.9$, the ratio of the CREs for the single-photon augmented case (blue hexagon markers in Fig.~\ref{fig:DVaugmentation_sims}(a)) and the un-augmented coherent-state case (black dashed plot in Fig.~\ref{fig:DVaugmentation_sims}(a)) is $0.2063/0.1443 \approx 1.43$. In Fig.~\ref{fig:DVaugmentation_sims}(b), we plot the change-detection latency $\tau$ as a function of the amount of squeezing (in dB) added per coherent-state pulse (for the squeezing-augmented transmitter). The purple and cyan markers correspond to simulation runs for the un-augmented case without and with BPSK modulation imprinted on the coherent state pulses, respectively. The black dashed line is the simulated change-detection latency for the discrete-variable (DV) augmented transmitter. The ratio of the change-detection latency is roughly $68/48 \approx 1.42$, which matches the aforementioned CRE ratio. The red and green markers in Fig.~\ref{fig:DVaugmentation_sims}(b) correspond to random simulation runs for the squeezing-augmented case without and with BPSK modulation on the coherent amplitudes of the pulses, respectively. Change detection latency $\tau$ decreases with increasing squeezing, as expected. For the continuous-variable (CV) squeezing-augmented transmitter's performance to match the DV single-photon-augmented transmitter's performance, the squeezing value needed $r_{\rm th} \approx 0.21$, corresponding to roughly $1.83$ dB of squeezing per pulse. Even though the mean quantum photons per pulse for this squeezing value, $N_a \approx 0.045$, which is much lower than the $N_a = 1$ used by the DV-augmented transmitter to achieve the same CRE, one could argue that generating heralded single photons using a bank of switched spontaneous parametric downconversion sources and single photon detectors could be technologically easier than generating displaced coherent states with $1.83$ dB effective {\em pulsed} squeezing. On the other hand, technology pertaining to generating pulsed squeezed light is progressing rapidly, and with progressively more squeezing per pulse, the CRE is seen to drop for the CV-augmented case as the squeezing increases. Whether there are more advanced DV (or generally, other non-Gaussian) augmentation schemes that outperforms the single-photon augmentation of coherent states that we considered here, is left open for future research.

\section{Change detection in the absence of classical pulses: CV versus DV comparison}\label{app:CVDV}
In this section, we will consider the scenario when there are no strong coherent state pulses on which the quantum photons would ride, i.e., $N=0$ and $N_a > 0$ in our notation. We will begin with an analysis wherein loss-change detection is the only task at hand, i.e., there is no communications taking place simultaneously on the channel. Thereafter, we will consider a natural extension to the case where loss-change detection happens in conjunction with {\em quantum} communications. This is in contrast with what was discussed in the paper on quantum-augmented BPSK-modulated coherent-state transmission, where loss-change detection happened in conjunction with {\em classical} communications.

\subsection{Loss-change detection with single-photon probe}

Let us consider a pure-loss channel of initial transmissivity $\eta_1$, and post-change transmissivity $\eta_2 = \eta_1\, \eta_{\rm tap}$ with $\eta_{\rm tap} = 0.9$. We will consider four transmitter probes, each transmitting $1$-mean-photon-number pulses:
\begin{itemize}
\item {\textbf{Coherent state}}: Pulses of weak coherent states $|\alpha\rangle$, with $|\alpha|^2 = 1$, $\alpha \in {\mathbb R}$. We will pair this transmitter with: (1) homodyne-detection receiver, and (2) Kennedy receiver with post-nulling residual photons per pulse $N_\epsilon$, where we will evaluate its performance for $N_\epsilon = 10^{-4}$ and $N_\epsilon = 10^{-5}$.
\item {\textbf{Squeezed-vacuum state}}: Pulses of squeezed-vacuum states $|0;r\rangle$, with sinh${}^2(r) = 1$. We will pair this transmitter probe with a homodyne detection receiver at the channel output.
\item {\textbf{Entangled squeezed state}}: Blocks of $n$ entangled pulses generated by splitting a squeezed-vacuum state $|0;s\rangle$, with sinh${}^2(s) = n$, in a linear uniform $n$-splitter such as the $n$-mode Green Machine~\cite{Cui2023}. We will pair this with a homodyne detection receiver, detecting each pulse independently (thereby generating a sequence of i.i.d. $n$-variate-correlated Gaussian random-vectors, as described in the paper earlier).
\item {\textbf{Single-photon Fock state}}: Pulses of single photon states $|1\rangle$. We will pair this transmitter with: (1) homodyne-detection receiver, and (2) direct detection (single-photon-detection) receiver. 
\end{itemize}

\begin{figure}[t]
\centering
\includegraphics[width=\columnwidth]{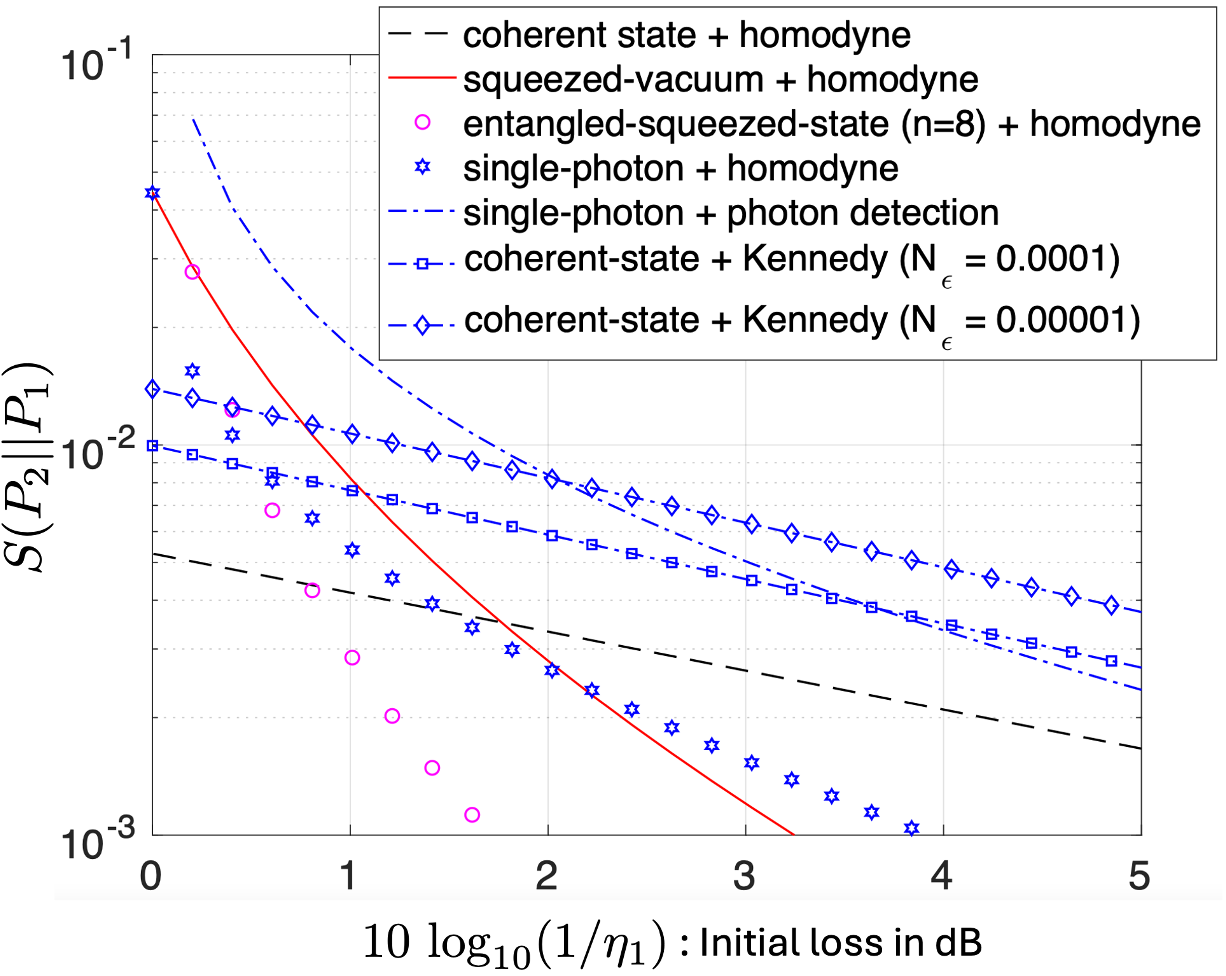}
\caption{A small-signal quantum pulse train---one-photon per pulse---is employed for loss-change detection. In the notation of the paper, $N=0$ and $N_a=1$. We plot the CRE $S(P_2||P_1)$ as a function of the pre-change loss in dB. We assume $\eta_{\rm tap} = 0.9$ for these plots.}
\label{fig:CVDV_comparison}
\end{figure}

We summarize the results in Fig.~\ref{fig:CVDV_comparison}, where we plot the CRE $S(P_2||P_1)$ as a function of the pre-change loss in dB, for all the aforesaid seven cases. First, let us recall from the previous section, Fig.~\ref{fig:DVaugmentation_sims}, that in the presence of strong coherent-state (e.g., classical-communications) pulses, i.e., $N \gg N_a$, the discrete-variable (DV) single-photon-augmented transmitter does worse compared to the CV-augmented cases. There was a small region of low pre-change loss where the CRE of the DV augmentation outperformed the un-augmented coherent-state case. In Fig.~\ref{fig:CVDV_comparison} we see however, that in the absence of base coherent-state pulses ($N=0, N_a = 1$), the trend is reversed, where the DV single-photon probe paired with a single-photon-detection receiver outperforms all the CV-probe cases. A single photon $|1\rangle$ transmitted through a lossy channel of transmissivity $\eta_1$ results in the (mixed) state $\eta_1 |1\rangle \langle 1| + (1-\eta_1)|0\rangle \langle 0 |$ at the channel, which is diagonal in the Fock basis, making a Fock-basis measurement (i.e., photon number detection) the optimum (CRE-maximizing) measurement. Homodyne detection on the other hand is a far-inferior measurement choice here. A coherent state is the only quantum state that retains its purity through lossy transmission, i.e., the transmission of the coherent state probe $|\alpha \rangle$, $|\alpha|^2 = 1$ results in a coherent state $|\sqrt{\eta_1}\alpha \rangle$, with mean photon number $|\sqrt{\eta_1}\alpha|^2 = \eta_1$ at the channel output. Because the output state is pure pre-change and post-change, the QRE is infinity, and for the same reason as described in the previous section, the CRE achieved by the Kennedy receiver~\cite{Kennedy1973} is infinity. As before however, the Kennedy receiver's CRE is extremely sensitive to the residual photons post-nulling (of the pre-change output state), $N_\epsilon$, as seen in Fig.~\ref{fig:CVDV_comparison}, making the simple single-photon probe paired with single-photon detection the best choice overall, for any initial channel loss~\footnote{Perfect nulling is hard to achieve by the Kennedy receiver since it requires not only acquiring a phase reference to the received pulses (much like homodyne detection does, e.g., using an optical phase-locked loop), but it also requires obtaining a precise amplitude reference to the received pulses. Further, the intuition behind why the Kennedy receiver gives a performance boost over homodyne detection even in the absence of perfect nulling is similar to why the Kennedy receiver is seen to perform well even for noisy coherent states~\cite{Jagannathan2022} and that over-nulling Kennedy-style receivers do better both for BPSK state discrimination~\cite{Takeoka2008} as well as for demodulating pulse position modulation (PPM)~\cite{Guha2010,Chen2012}.}. 

\subsection{Joint quantum communications and loss-change detection with single-photon transmitter}~\label{sec:jointcommsensing}

We have seen that the quantum-augmentation strategies presented in the Letter barely affect the channel's classical-communications capacity, while providing a sharp boost to the smallest latency with which a change (in the channel loss) can be detected. A few lines of enquiry are almost obvious. One is the general question of what is the optimal {\em tradeoff region} for jointly executing the two tasks (of communications and change detection), quantified in terms of the tradeoff between communications capacity and change-detection latency (of a channel characteristic such as transmission loss or thermal noise). The transmitter's optimal choice of the ensemble and distribution of quantum states to encode information, versus the optimal choice of the ensemble and distribution of quantum states at the transmitter for minimum-latency change-detection may not be the same. It was recently shown, for the setting of the classical AWGN channel, that one can outperform the strategy of time-sharing between the respective-optimal strategies for the two tasks~\cite{Seo2024}. Evaluating the fundamental information-theoretic tradeoff region of communications capacity and change-detection of loss, over a bosonic channel, and the development of associated modulation, coding and receiver strategies to achieve that fundamental region, is a topic being left open for future research. 

The second obvious line of enquiry that emerges is whether the ideas presented in the Letter generalize to channel change detection performed along side {\em quantum communications}. Arguably the simplest quantum modulation format---way to encode quantum information in optical pulses---is the time-bin {\em dual-rail qubit} that encodes one qubit into two pulses of light containing exactly one photon. The state $|1\rangle |0\rangle$ containing one photon in the first time bin and vacuum in the second time bin encodes the logical $|0\rangle_L$ state of the qubit, whereas $|0\rangle |1\rangle \equiv |1\rangle_L$. Any other state of the qubit $c_0 |0\rangle_L + c_1 |1\rangle_L$, $|c_0|^2 + |c_1|^2 = 1$, can be prepared by mixing the two time bins of $|0\rangle_L$ (or $|1\rangle_L$) in a two-input two-output beamsplitter (after delaying the first time bin appropriately). This implies that any single-qubit gate (unitary) can be realized using a single beamsplitter. Measurement in the $Z$ (computational) basis, i.e., the projective measurement $\left\{|0\rangle_L {}_L\langle 0|, |1\rangle_L {}_L\langle 1|\right\}$, can be performed by single photon detection on both time bins carrying the qubit. Therefore, a single-qubit measurement in any basis can be performed by applying a beamsplitter followed by single-photon detection. The most well-known protocol to transfer the state of a time-bin dual-rail qubit into an atomic qubit (such as a color center, trapped ion, or neutral atom qubit, etc.) is the Duan-Kimble scheme~\cite{DuanKimble2004}. Two salient features of this scheme are that: (a) it deterministically transfers the state of a dual-rail photonic qubit into the atomic qubit, and (b) if the dual rail photonic qubit had undergone (pure) loss prior to the attempted state transfer into the quantum memory, it would herald the loss of the photon in transit. In other words, if neither of the two photon detectors of the Duan Kimble scheme clicks, one would know that the photonic qubit had failed to make it through the lossy channel. On the other hand, if one of the two detectors clicks, we would know (for certain) that the state of the photonic qubit was transferred successfully to the atomic qubit. Depending upon which of the two detectors clicked, one has to apply a phase correction (or not) to the state of the atomic qubit, to ensure this state transfer to work.

The term {\em quantum communications} is used inter-changeably to refer to either: (1) quantum key distribution (QKD)---a protocol suite that generates shared keys between distant parties whose security underlies laws of quantum physics as opposed to assumptions on the computational power of the adversary; or to: (2) faithful transmission of qubits from one point to another, e.g., between two quantum computers or from one quantum repeater to another one across the link of a quantum network. If one were executing QKD using the BB84 protocol~\cite{Bennett2014} (task 1) using dual-rail qubits, the receiver would use a single photon detector. So, it would herald the successful arrival of the photonic qubit over the lossy channel, and hence would be able to weed out only those instances where the qubit made it to the channel's output. If one were transmitting qubits from one quantum memory register to another (task 2) using dual-rail qubits, the Duan-Kimble gadget at the receiver would herald the successful arrival of the photonic qubit over the lossy channel, and hence would be able to weed out only those instances where the transmitted qubit did make it successfully to the channel output. 

The CRE computation shown in Fig.~\ref{fig:CVDV_comparison} for the single-photon transmitter paired with single-photon detection is based on the pre-change distribution: $P_1[0] = 1-\eta_1, P_1[1] = \eta_1$, and post-change distribution: $P_2[0] = 1-\eta_2, P_2[1] = \eta_2$. The CUSUM algorithm for change detection would use instances of the binary-valued random variable with the above distribution. It is simple to see that---regardless of whether the communicating parties were engaged in QKD or a qubit-communications session---that the detector's (or, respectively the Duan-Kimble gadget's) output is a classical random variable with exactly the above distribution. This is because any modulation of the qubit at the transmitter does not alter the above distribution, since the distribution only relies on the probability of the photon carrying the dual-rail qubit surviving the channel loss (or not). Therefore, one can run the CUSUM algorithm for change detection without affecting the rate or fidelity of the underlying quantum communications session.

The performance tradeoff for other photonic quantum modulation formats (such as the square-GKP, hex-GKP, cat-state, etc.), and the ultimate trade-space of joint quantum-communications and change-detection is left open for future research.

\section{Numerical computation of the Average Run Length (ARL)}\label{app:ARL}
To detect a change in transmissivity from $\eta_1$ to $\eta_2$ -- and hence from the pre-change probability distribution, $P_1$, to the post-change distribution, $P_2$ -- the CUSUM algorithm sequentially accumulates a sum of Log Likelihood Ratios (LLRs) 
$\sum_{n=1}^k \log[P_2(X_n)/P_1(X_n)]$ and the decision function $G[k] = \max_{1 \le n_c \le k}\sum_{n=n_c}^k \log[P_2(X_n)/P_1(X_n)]$,
and calculates the smallest $k=k^*$ for which $G[k] > h$, where $h$ is a user-defined confidence measure. If there is no change, the random variates $\{X_n\}_{n\geq 1}$ will continue to obey $P_1$ and $G[k^*]$ would eventually exceed $h$ thus causing a false alarm. $k^* = \gamma(h)$, which is the average time to raise a false alarm, is called the \textit{average run length} (ARL). 

For any given pre-change distribution $P_1$, we compute $\gamma(h)$ for a $L$ finely sampled values $h_j \in [h_{\rm min},h_{\rm max}]$ using a Monte Carlo method as follows (we assume that $h_j$'s are monotonically increasing with $j$ and are equally spaced between $h_{\rm min}$ and $h_{\rm max}$). 
In each independent run, we draw $N$ samples $\{X_n\}_{n=1}^N$ at random from $P_1$ where $N$ is significantly greater than the likely time to false alarm. We initialize $j=1$ and compute $G[k]$ for each $k$ such that $0\leq k \leq N$. If $G[k] > h_j$, we save $\gamma(h_j) = k$ and increment $j$; otherwise we increment $k$. At the end of run $i$, we have a table $\gamma_i = \{\gamma(h_1=h_{\rm min}),\gamma(h_2),\gamma(h_3),\ldots,\gamma(h_L=h_{\rm max})\}$. Since $\gamma_i$ is a monotonically non-decreasing sequence of length $L$, we average $M$ independent runs and compute the average vector $\gamma = \frac{1}{M}\sum_{i=1}^M \gamma_i$, which yields accurate estimates of ARL at various values of $h$. Since ARL increases monotonically with $h$, our method has significantly lower variance compared to running the procedure independently for each $h\in [h_{\rm min},h_{\rm max}]$ and then averaging.

During the computation of ARL, drawing random samples $X_n$ is an easy task for the CV case since both $P_1$ and $P_2$ are Gaussian distributions (with different means and variances). However, the DV distributions (Eq. \ref{eq:pdfDV}) are non-Gaussian and bimodal, therefore we draw a random sample $x^*$ by first sampling a uniform random number $u \sim U([0,1])$ and then numerically computing the real root $x^*$ of $F(x)=u$, where $F(x)$ is the CDF of the DV distribution in Eq. \ref{eq:pdfDV} as given below:
\begin{align}\label{eq:cdfDV}
    F(x)&:= \sqrt{\frac{2}{\pi }} e^{-2 \left(x-\alpha\sqrt{\eta }\right)^2} (-x + \alpha\sqrt{\eta}) \eta \:+\nonumber \\
    & \frac{1}{2}(1+{\rm erf}(\sqrt{2}\left(x-\alpha\sqrt{\eta }\right))).
\end{align}

For the plot in Figure \ref{fig:DVaugmentation_sims}(b), we fix ARL to $2\times 10^6$ time slots (or pulses) for all schemes compared. For each scheme $s$ considered, we look up the value of $h_s$ in the table $\gamma_s$ such that $\gamma_s(h_s) = 2\times 10^6$ and use $h_s$ as an input to the CUSUM algorithm to compute detection latency.


\begin{thebibliography}{10}
\providecommand{\url}[1]{#1}
\csname url@samestyle\endcsname
\providecommand{\newblock}{\relax}
\providecommand{\bibinfo}[2]{#2}
\providecommand{\BIBentrySTDinterwordspacing}{\spaceskip=0pt\relax}
\providecommand{\BIBentryALTinterwordstretchfactor}{4}
\providecommand{\BIBentryALTinterwordspacing}{\spaceskip=\fontdimen2\font plus
\BIBentryALTinterwordstretchfactor\fontdimen3\font minus
  \fontdimen4\font\relax}
\providecommand{\BIBforeignlanguage}[2]{{%
\expandafter\ifx\csname l@#1\endcsname\relax
\typeout{** WARNING: IEEEtran.bst: No hyphenation pattern has been}%
\typeout{** loaded for the language `#1'. Using the pattern for}%
\typeout{** the default language instead.}%
\else
\language=\csname l@#1\endcsname
\fi
#2}}
\providecommand{\BIBdecl}{\relax}
\BIBdecl

\bibitem{Shim2012}
H.~K. Shim, K.~Y. Cho, Y.~Takushima, and Y.~C. Chung,
  ``\BIBforeignlanguage{en}{Correlation-based {OTDR} for in-service monitoring
  of 64-split {TDM} {PON}},'' \emph{\BIBforeignlanguage{en}{Opt. Express}},
  vol.~20, no.~5, pp. 4921--4926, Feb. 2012.

\bibitem{Iqbal2011}
M.~Zafar~Iqbal, H.~Fathallah, and N.~Belhadj, ``\BIBforeignlanguage{en}{Optical
  fiber tapping: Methods and precautions},'' in
  \emph{\BIBforeignlanguage{en}{8th International Conference on High-capacity
  Optical Networks and Emerging Technologies}}.\hskip 1em plus 0.5em minus
  0.4em\relax IEEE, Dec. 2011, pp. 164--168.

\bibitem{Fok2011}
M.~P. Fok, Z.~Wang, Y.~Deng, and P.~R. Prucnal,
  ``\BIBforeignlanguage{en}{Optical layer security in fiber-optic networks},''
  \emph{\BIBforeignlanguage{en}{IEEE Trans. Inf. Forensics Secur.}}, vol.~6,
  no.~3, pp. 725--736, Sep. 2011.

\bibitem{Medard2002}
M.~Medard, S.~R. Chinn, and P.~Saengudomlert, ``\BIBforeignlanguage{en}{Attack
  detection in all-optical networks},'' in \emph{\BIBforeignlanguage{en}{OFC
  '98. Optical Fiber Communication Conference and Exhibit. Technical Digest.
  Conference Edition. 1998 OSA Technical Digest Series Vol.2 (IEEE Cat.
  No.98CH36177)}}.\hskip 1em plus 0.5em minus 0.4em\relax Opt. Soc. America,
  2002, pp. 272--273.

\bibitem{Gong2020}
Y.~Gong, R.~Kumar, A.~Wonfor, S.~Ren, R.~V. Penty, and I.~H. White,
  ``\BIBforeignlanguage{en}{Secure optical communication using a quantum
  alarm},'' \emph{\BIBforeignlanguage{en}{Light Sci. Appl.}}, vol.~9, no.~1, p.
  170, Oct. 2020.

\bibitem{Monras2007-jg}
\BIBentryALTinterwordspacing
A.~Monras and M.~G.~A. Paris, ``\BIBforeignlanguage{en}{{Optimal quantum
  estimation of loss in bosonic channels}},''
  \emph{\BIBforeignlanguage{en}{Phys. Rev. Lett.}}, vol.~98, no.~16, p. 160401,
  Apr. 2007. [Online]. Available:
  \url{http://dx.doi.org/10.1103/PhysRevLett.98.160401}
\BIBentrySTDinterwordspacing

\bibitem{Gong2023}
Z.~Gong, N.~Rodriguez, C.~N. Gagatsos, S.~Guha, and B.~A. Bash,
  ``{Quantum-Enhanced} transmittance sensing,'' \emph{IEEE J. Sel. Top. Signal
  Process.}, vol.~17, no.~2, pp. 473--490, Mar. 2023.

\bibitem{Pirandola2011-xo}
\BIBentryALTinterwordspacing
S.~Pirandola, ``\BIBforeignlanguage{en}{{Quantum reading of a classical digital
  memory}},'' \emph{\BIBforeignlanguage{en}{Phys. Rev. Lett.}}, vol. 106,
  no.~9, p. 090504, Mar. 2011. [Online]. Available:
  \url{http://dx.doi.org/10.1103/PhysRevLett.106.090504}
\BIBentrySTDinterwordspacing

\bibitem{Shapiro1979}
J.~Shapiro, H.~Yuen, and A.~Mata, ``Optical communication with two-photon
  coherent states--part {II}: Photoemissive detection and structured receiver
  performance,'' \emph{IEEE Trans. Inf. Theory}, vol.~25, no.~2, pp. 179--192,
  Mar. 1979.

\bibitem{Cui2023}
C.~Cui, J.~Postlewaite, B.~N. Saif, L.~Fan, and S.~Guha,
  ``\BIBforeignlanguage{en}{Superadditive communication with the green machine
  as a practical demonstration of nonlocality without entanglement},''
  \emph{\BIBforeignlanguage{en}{Nat. Commun.}}, vol.~16, no.~1, p. 3760, Apr.
  2025.

\bibitem{Page1957}
E.~S. Page, ``On problems in which a change in a parameter occurs at an unknown
  point,'' \emph{Biometrika}, vol.~44, no. 1-2, pp. 248--252, Jun. 1957.

\bibitem{Lorden1971}
G.~Lorden, ``Procedures for reacting to a change in distribution,'' \emph{Ann.
  Math. Stat.}, vol.~42, no.~6, pp. 1897--1908, 1971.

\bibitem{Sentis2017}
G.~Sentís, J.~Calsamiglia, and R.~Muñoz-Tapia,
  ``\BIBforeignlanguage{en}{Exact identification of a quantum change point},''
  \emph{\BIBforeignlanguage{en}{Phys. Rev. Lett.}}, vol. 119, no.~14, p.
  140506, Oct. 2017.

\bibitem{Sentis2018}
G.~Sentís, E.~Martínez-Vargas, and R.~Muñoz-Tapia,
  ``\BIBforeignlanguage{en}{Online strategies for exactly identifying a quantum
  change point},'' \emph{\BIBforeignlanguage{en}{Phys. Rev. A (Coll. Park.)}},
  vol.~98, no.~5, p. 052305, Nov. 2018.

\bibitem{Fanizza2022-tu}
\BIBentryALTinterwordspacing
M.~Fanizza, C.~Hirche, and J.~Calsamiglia, ``{QUSUM: quickest quantum
  change-point detection},'' Aug. 2022. [Online]. Available:
  \url{http://arxiv.org/abs/2208.03265}
\BIBentrySTDinterwordspacing

\bibitem{Fanizza2023-ch}
\BIBentryALTinterwordspacing
------, ``\BIBforeignlanguage{en}{{Ultimate Limits for Quickest Quantum
  Change-Point Detection}},'' \emph{\BIBforeignlanguage{en}{Phys. Rev. Lett.}},
  vol. 131, no.~2, p. 020602, Jul. 2023. [Online]. Available:
  \url{http://dx.doi.org/10.1103/PhysRevLett.131.020602}
\BIBentrySTDinterwordspacing

\bibitem{John2025-bz}
\BIBentryALTinterwordspacing
T.~C. John, C.~N. Gagatsos, and B.~A. Bash, ``{Fundamental limits of quickest
  change-point detection with continuous-variable quantum states},''
  \emph{arXiv [quant-ph]}, Apr. 2025. [Online]. Available:
  \url{http://arxiv.org/abs/2504.16259}
\BIBentrySTDinterwordspacing

\bibitem{Yu2018}
S.~Yu, C.-J. Huang, J.-S. Tang, Z.-A. Jia, Y.-T. Wang, Z.-J. Ke, W.~Liu,
  X.~Liu, Z.-Q. Zhou, Z.-D. Cheng, J.-S. Xu, Y.-C. Wu, Y.-Y. Zhao, G.-Y. Xiang,
  C.-F. Li, G.-C. Guo, G.~Sentís, and R.~Muñoz-Tapia,
  ``\BIBforeignlanguage{en}{Experimentally detecting a quantum change point via
  the bayesian inference},'' \emph{\BIBforeignlanguage{en}{Phys. Rev. A}},
  vol.~98, no.~4, p. 040301, Oct. 2018.

\bibitem{Note1}
Incorporating excess noise, such as stemming from electronic noise or local
  oscillator intensity fluctuations causing excess noise due to imperfect
  common mode rejection ratio (CMRR) of homodyne detection, results in
  qualitative changes to results presented herein, but are not important for
  the main message of the Letter.

\bibitem{Goel1971}
\BIBentryALTinterwordspacing
A.~L. Goel and S.~M. Wu, ``Determination of a. r. l. and a contour nomogram for
  cusum charts to control normal mean,'' \emph{Technometrics}, vol.~13, no.~2,
  pp. 221--230, 1971. [Online]. Available:
  \url{http://www.jstor.org/stable/1266785}
\BIBentrySTDinterwordspacing

\bibitem{SM}
S.~Guha, T.~Cherian~John, Z.~Gong, and P.~Basu, ``Supplemental materials:
  Quantum-enhanced quickest change detection of transmission loss,''
  \emph{Phys. Rev. Lett. (submitted: supplemental materials)}, 2025.

\bibitem{Guha2011}
S.~Guha, ``Structured optical receivers to attain superadditive capacity and
  the holevo limit,'' \emph{Phys. Rev. Lett.}, 2011.

\bibitem{Hamilton2017}
C.~S. Hamilton, R.~Kruse, L.~Sansoni, S.~Barkhofen, C.~Silberhorn, and I.~Jex,
  ``\BIBforeignlanguage{en}{Gaussian boson sampling},''
  \emph{\BIBforeignlanguage{en}{Phys. Rev. Lett.}}, vol. 119, no.~17, p.
  170501, Oct. 2017.

\bibitem{Huettinger2006}
S.~Huettinger and J.~Huber, ``Analysis and design of power-efficient coding
  schemes with parallel concatenated convolutional codes,'' \emph{IEEE Trans.
  Commun.}, vol.~54, no.~7, pp. 1251--1258, Jul. 2006.

\bibitem{Kennedy1973}
R.~S. Kennedy, ``A near-optimum receiver for the binary coherent state quantum
  channel,'' \emph{Research Laboratory of Electronics, MIT, Quarterly Progress
  Report}, vol. 108, pp. 219--225, 1973.

\bibitem{Jagannathan2022}
A.~Jagannathan, M.~Grace, O.~Brasher, J.~H. Shapiro, S.~Guha, and J.~L. Habif,
  ``Demonstration of quantum-limited discrimination of multicopy pure versus
  mixed states,'' \emph{Phys. Rev. A}, vol. 105, no.~3, p. 032446, Mar. 2022.

\bibitem{Seo2024}
D.~Seo and S.~H. Lim, ``On the fundamental tradeoff of joint communication and
  quickest change detection,'' Jan. 2024.

\bibitem{DuanKimble2004}
\BIBentryALTinterwordspacing
L.-M. Duan and H.~J. Kimble, ``Scalable photonic quantum computation through
  cavity-assisted interactions,'' \emph{Phys. Rev. Lett.}, vol.~92, p. 127902,
  Mar 2004. [Online]. Available:
  \url{https://link.aps.org/doi/10.1103/PhysRevLett.92.127902}
\BIBentrySTDinterwordspacing

\bibitem{Bennett2014}
C.~H. Bennett and G.~Brassard, ``\BIBforeignlanguage{en}{Quantum cryptography:
  Public key distribution and coin tossing},''
  \emph{\BIBforeignlanguage{en}{Theor. Comput. Sci.}}, vol. 560, pp. 7--11,
  Dec. 2014.

\bibitem{Takeoka2008}
M.~Takeoka and M.~Sasaki, ``Discrimination of the binary coherent signal:
  Gaussian-operation limit and simple non-gaussian near-optimal receivers,''
  \emph{Phys. Rev. A}, vol.~78, no.~2, p. 022320, Aug. 2008.

\bibitem{Guha2010}
S.~Guha, J.~L. Habif, and M.~Takeoka, ``{PPM} demodulation: On approaching
  fundamental limits of optical communications,'' in \emph{2010 {IEEE}
  International Symposium on Information Theory}.\hskip 1em plus 0.5em minus
  0.4em\relax IEEE, Jun. 2010, pp. 2038--2042.

\bibitem{Chen2012}
J.~Chen, J.~L. Habif, Z.~Dutton, R.~Lazarus, and S.~Guha,
  ``\BIBforeignlanguage{en}{Optical codeword demodulation with error rates
  below the standard quantum limit using a conditional nulling receiver},''
  \emph{\BIBforeignlanguage{en}{Nat. Photonics}}, vol.~6, no.~6, pp. 374--379,
  May 2012.

\bibitem{Gong2025}
Z.~Gong and S.~Guha, ``\BIBforeignlanguage{en}{Quantum-enhanced change
  detection and joint communication detection},''
  \emph{\BIBforeignlanguage{en}{Physical Review A}}, vol. 112, no.~3, p.
  032604, Sep. 2025.

\bibitem{Wilde2012}
M.~M. Wilde, P.~Hayden, and S.~Guha, ``\BIBforeignlanguage{en}{Information
  trade-offs for optical quantum communication},''
  \emph{\BIBforeignlanguage{en}{Phys. Rev. Lett.}}, vol. 108, no.~14, p.
  140501, Apr. 2012.

\bibitem{Note2}
Note that in the convention used in this paper the Fourier transform of a
  distribution $\mu $ on the real line is given by integrating the function
  $e^{-itx}$ with respect to $\mu (dx)$.

\bibitem{Note3}
Perfect nulling is hard to achieve by the Kennedy receiver since it requires
  not only acquiring a phase reference to the received pulses (much like
  homodyne detection does, e.g., using an optical phase-locked loop), but it
  also requires obtaining a precise amplitude reference to the received pulses.
  Further, the intuition behind why the Kennedy receiver gives a performance
  boost over homodyne detection even in the absence of perfect nulling is
  similar to why the Kennedy receiver is seen to perform well even for noisy
  coherent states~\cite {Jagannathan2022} and that over-nulling Kennedy-style
  receivers do better both for BPSK state discrimination~\cite {Takeoka2008} as
  well as for demodulating pulse position modulation (PPM)~\cite
  {Guha2010,Chen2012}.
  
   \bibitem{DataSet}
  Supplemental dataset for Figures 3 and 7, \url{https://doi.org/10.5281/zenodo.17460928}.

\end{thebibliography}
\end{document}